\documentclass[epj,final]{svjour}

\pdfoutput=1

\usepackage{bm}
\usepackage[pdftex]{graphicx}
\usepackage{amssymb}
\def\z{{\hat{\bm z}}}

\def\r{{\bm r}}
\def\R{{\bm R}}

\def\q{{\bf q}}
\def\p{{\bf p}}
\def\k{{\bf k}}

\def\0{{\bf 0}}
\def\Q{{\bf Q}}

\def\down{\downarrow}
\def\up{\uparrow}
\def\nn{\nonumber}

\def\rf#1{(\ref{#1})}
\def\rfs#1{Eq.~\rf{#1}}

\begin{document}
\title{Atom-dimer and dimer-dimer scattering in fermionic mixtures
  near a narrow Feshbach resonance}
\author{J.~Levinsen\inst{1,2} \and D.~S.~Petrov\inst{2,3}}
\institute{
T.C.M. Group, University of Cambridge, Cavendish Laboratory,
J. J. Thomson Ave., Cambridge CB3
0HE, United Kingdom
 \and
Laboratoire de
  Physique Th\'eorique et Mod\`eles Statistiques, CNRS and Universit\'e
  Paris Sud, UMR8626, B\^at. 100, 91405 Orsay, France
 \and
 Russian Research Center Kurchatov Institute, Kurchatov
  Square, 123182 Moscow, Russia}
\date{Received: date / Revised version: date}
\abstract{
  We develop a diagrammatic approach for solving few-body problems in
  heteronuclear fermionic mixtures near a narrow interspecies Feshbach
  resonance. We calculate $s$-, $p$-, and $d$-wave phaseshifts for the
  scattering of an atom by a weakly-bound dimer. The fermionic
  statistics of atoms and the composite nature of the dimer lead to a
  strong angular momentum dependence of the atom-dimer interaction,
  which manifests itself in a peculiar interference of the scattered
  $s$- and $p$-waves. This effect strengthens with the mass ratio and
  is remarkably pronounced in $^{40}$K-($^{40}$K-$^{6}$Li) atom-dimer
  collisions. We calculate the scattering length for two dimers formed near a narrow
  interspecies resonance. Finally, we discuss the collisional relaxation of the dimers to
  deeply bound states and evaluate the corresponding rate constant as
  a function of the detuning and collision energy.}

\authorrunning{J. Levinsen and D. S. Petrov}
\titlerunning{Scattering in fermionic mixtures near a narrow Feshbach resonance}

\maketitle

\section{Introduction}

Numerous advances in the field of ultracold Fermi gases over the past
decade have enabled the exploration of novel strongly interacting
regimes in fermionic systems (see \cite{Inguscio2008} and
\cite{Giorgini2008} for review). The BCS-BEC crossover, extensively
studied in either Potassium ($^{40}$K) or Lithium ($^6$Li) homonuclear
systems, is now being actively pursued in the new generation of
experiments on mixtures of these two isotopes
\cite{Taglieber2008,Wille2008,Voigt2009,Spiegelhalder2009,Tiecke2009,Spiegelhalder2010,Dieckmann2010,Trenkwalder2010}.
The mass ratio is thus a new parameter introduced into the crossover
phase diagram \cite{note:optlat}. It is then natural to ask whether
changing this parameter can lead to qualitatively new crossover
physics and, if so, how large should the mass ratio be in order to see
non-trivial effects?
%Can it lead to
%qualitatively new crossover physics and how large should be the mass
%ratio in order to see non-trivial effects?

For a very large mass ratio (of the order of several hundreds) a
crystalline phase can emerge on the BEC side of the crossover
\cite{Petrov2007}. The effect is due to a strong long-range repulsion
between the heavy fermions originating from the exchange of their
light partners. Another mass ratio dependent change in the behavior of
the system occurs in the problem of two identical heavy fermions of
mass $m_\up$ interacting resonantly with a light atom of mass
$m_\down$. The exchange of the light atom results in an attractive
potential between the heavy fermions proportional to $1/m_\down R^2$,
where $R$ is the distance between them. This exchange attraction
competes with the repulsive centrifugal barrier $\propto 1/m_\up R^2$
for the identical fermions. For mass ratios $m_\up/m_\down$ larger
than the critical value 13.6 the exchange attraction dominates over
the centrifugal barrier and the heavy particles experience a fall to
the centre in the $1/R^2$-potential \cite{LL}. On resonance
(interspecies scattering length $a=\infty$) this leads to the Efimov
effect - the existence of an infinite number of bound
heavy-heavy-light trimer states
\cite{Efimov1973,Fonseca1979,Amado1971}. For smaller mass ratios the
centrifugal barrier is dominant. On the one hand this effective
three-body repulsion excludes the Efimov effect. On the other it
suppresses recombination processes requiring three atoms to approach
each other to very short distances, which is evidently advantageous
for the collisional stability of the gas. The lower the mass ratio,
the more stable this three-body system is \cite{Petrov2005a}.

It turns out that even for $m_\up/m_\down<13.6$ the
$\up\up\down$-system exhibits non-perturbative effects on the positive
(BEC) side of the resonance, where there is a weakly bound
heteronuclear molecular state. One of us found that the three-body
recombination to this state vanishes for $m_\up/m_\down\approx 8.6$
\cite{Petrov2003}. Later, Kartavtsev and Malykh argued that this
phenomenon is related to the existence of a weakly bound {\it not}
Efimovian trimer state for $m_\up/m_\down>8.2$
\cite{Kartavtsev2007}. The trimer has unit angular momentum and for
smaller mass ratios turns into a $p$-wave atom-dimer scattering
resonance. We have recently shown that in the case of a K-Li mixture
($m_\up/m_\down=6.64$) the K-(K-Li) atom-dimer scattering should be
dominated by this $p$-wave resonance in a wide range of collision
energies \cite{Levinsen2009}. Moreover, by introducing an external
quasi-2D confinement, the $p$-wave atom-dimer interaction can be tuned
from attractive to repulsive, allowing for a trimer formation.

In this paper we develop a uniform-space diagrammatic approach for
studying few-body processes in a heteronuclear fermionic mixture near
an interspecies Feshbach resonance of finite width. We calculate
relevant atom-dimer scattering phaseshifts and partial cross-sections
in the homonuclear case and in the K-Li case. Passing from
$m_\up/m_\down=1$ to $m_\up/m_\down=6.64$ we observe an increase in
the atom-dimer interaction, repulsive in even angular momentum
channels and attractive in odd ones. The most dramatic increase is
found in the channel with unit angular momentum -- the $p$-wave
scattering volume changes by more than an order of magnitude. Our
exact calculations are complemented by a qualitative explanation of
the observed effect based on the Born-Oppenheimer approach, which we
generalize to the case of a narrow interspecies resonance. We predict
a very strong interference between $s$- and $p$-waves in atom-dimer
scattering. Depending on the collision energy, the scattering is
dominant in backward or forward directions, which can be observed
experimentally by colliding an atomic cloud with a cloud of
molecules. We use our diagrammatic approach to calculate the
dimer-dimer scattering length $a_{dd}$ as a function of the atomic
scattering length $a$ and the width of the interspecies
resonance. Finally, we discuss the main mechanisms of the collisional
relaxation of dimers into deep molecular states, and calculate the
corresponding atom-dimer and dimer-dimer relaxation rate constants as
functions of $a$, the width of the resonance, and the collision
energy.

The paper is organized as follows. In Sec.~\ref{sec:exact} we discuss
the two-body problem in the narrow resonance case and introduce our
field-theoretical approach. The main part of the paper is structured
according to the previous paragraph -- Secs. \ref{sec:3body} and
\ref{sec:4body} are devoted to the three- and four-body problems
respectively. In Sec.~\ref{sec:relaxation} we discuss the inelastic
collisional relaxation in atom-dimer and dimer-dimer collisions, and
in Sec.~\ref{sec:conc} we conclude.

\section{Two-body problem near a narrow Feshbach resonance \label{sec:exact}}

We assume that all interatomic interactions in the $\up$-$\down$ fermionic mixture are characterized by van der Waals potentials. We also assume that the intraspecies interactions are not resonant, and therefore can be safely neglected in the ultracold regime. Let us denote the van der Waals range of the interspecies interaction by $R_e$ and write down the partial wave expansion of the on-shell scattering amplitude \cite{LL}
\begin{equation}
f(\k,\k') = \sum_{\ell=0}^\infty(2\ell+1)P_\ell(\cos\theta)f_\ell(k).
\label{eq:f}
\end{equation}
Here $\k$ and $\k'$ are initial and final relative momenta such that $|\k|=|\k'|=k$, and $\theta=\angle_{\k,\k'}$ is the scattering angle. We set $\hbar=1$. The partial wave amplitudes $f_\ell(k)$ can be written in terms of the phase shifts $\delta_\ell(k)$ as
\begin{equation}
f_\ell(k) = \frac1{k\cot\delta_\ell(k)-ik}.
\label{eq:fphase}
\end{equation}
Expanding the denominator of Eq.~(\ref{eq:fphase}) in powers of $kR_e$ gives the effective range expansion. In particular, in the $s$-wave channel ($\ell=0$) we have
\begin{equation}
k\cot\delta_0(k)\approx-a^{-1}+\frac12 r_0 k^2+\dots,
\label{eq:delta0lowk}
\end{equation}
and the corresponding expansion in the $p$-wave channel ($\ell=1$) reads
\begin{equation}
k^3\cot\delta_1(k)\approx-v^{-1}+\frac12 k_0 k^2+\dots,
\label{eq:delta1lowk}
\end{equation}
where $v$ is the $p$-wave scattering volume, and $k_0$ is a parameter analogous to the effective range.

The scale of the scattering length $a$, the effective range $r_0$, and
other expansion parameters in the higher order terms in
Eq.~(\ref{eq:delta0lowk}) are set by the length $R_e$ or its power of
suitable dimension, and the same holds for higher partial waves. For
$k\rightarrow 0$ the partial scattering amplitudes are proportional to
$(kR_e)^{2\ell}$. Thus, in the limit $kR_e\ll 1$ (ultracold regime)
the $s$-wave scattering amplitude, which equals $f_0(0)=-a$, is the
most important interaction parameter in the mixture.

Near a scattering resonance the scattering length can be modified and,
in particular, can take anomalously large values (i.e. $|a| \gg
R_e$). A magnetic Feshbach resonance occurs when the collision energy
of the two atoms is close to the energy of a quasidiscrete molecular
state in another hyperfine domain, which is called closed channel. The
tuning of the scattering amplitude is achieved by shifting the open
and closed channels with respect to each other in an external magnetic
field (hyperfine states corresponding to the open and closed channel
have different magnetic moments). The width of the resonance is
determined by the strength of the coupling between these two
channels. The narrower the resonance, the stronger the collision
energy dependence of the scattering amplitude, and, therefore, the
larger the effective range $r_0$. We call a resonance narrow, if
$|r_0|\gg R_e$ \cite{note:manybody}. In fact, near such a resonance
$r_0$ is necessarily negative and it is convenient to use another
length parameter \cite{Petrov2004a}
\begin{equation}
R^*=-r_0/2=\frac1{2\mu a_{\mbox{\tiny bg}}\mu_{\mbox{\tiny rel}}\Delta B},
\end{equation}
where $\mu = m_\up m_\down/(m_\up+m_\down)$ is the reduced mass,
$a_{\mbox{\tiny bg}}$ is the background scattering length,
$\mu_{\mbox{\tiny rel}}$ is the difference in the magnetic moments of
the closed and open channels, and $\Delta B$ is the magnetic width of
the Feshbach resonance. All $^6$Li-$^{40}$K interspecies resonances
discussed so far are characterized by $R^*\gtrsim 100$nm
\cite{Wille2008,Tiecke2009}, which is much larger than the van der
Waals range $R_e\approx2.2$nm.

One can imagine an interatomic potential for which the higher order
terms in Eq.~(\ref{eq:delta0lowk}) are also anomalously large. For
example, we can introduce one or several additional closed channels
with quasistationary states very close to the open-channel threshold
resulting in a rather exotic scattering amplitude
\cite{note:model}. However, in this paper we assume a more practical
and simple case in which the terms denoted by $\dots$ in
Eq.~(\ref{eq:delta0lowk}) vanish in the limit $kR_e\rightarrow 0$. We
will also assume that scattering with $\ell > 0$ is not resonant and
can be neglected in this limit. Then, substituting
Eq.~(\ref{eq:delta0lowk}) into Eq.~(\ref{eq:fphase}) we get the
well-known formula for the resonant scattering at a quasidiscrete
level \cite{LL} (written as a function of momentum rather than energy)
\begin{equation}
f({\bf k}) = -\frac1{1/a + R^*k^2 + ik}.
\label{eq:fphase0}
\end{equation}
The ratio $R^*/a$ measures the detuning from the resonance and we
distinguish the regime of small detuning, $R^*/a\ll 1$, and the regime
of intermediate detuning, $R^*/a \gg 1$. The properties of a few-body
system in these two limits are qualitatively different
\cite{Petrov2004a}.

In order to describe the $\up\down$ mixture near a narrow resonance we
use the two-channel Hamiltonian \cite{Timmermans1999}
\begin{eqnarray}
\hat H & = & \sum_{\k,\sigma=\up,\down}\frac{k^2}{2m_\sigma}\hat 
a^\dagger_{\k,\sigma}\hat 
a_{\k,\sigma}+\sum_{\p}\left(\omega_0+\frac{p^2}{2M}\right)
\hat b^\dagger_\p \hat b_\p
\nn \\ && \hspace{-8mm}+\sum_{\k,\p}\frac g{\sqrt V}
\left(\hat b^\dagger_\p\hat a_{\frac{\p}2+\k,\up}\hat a_{\frac\p2-\k,\down}+
\hat b_\p\hat a^\dagger_{\frac\p2-\k,\down}\hat a^\dagger_{\frac\p2+\k,\up}\right),
\label{eq:hamiltonian}
\end{eqnarray}
where $a^\dagger_{\up,\down}$ and $a_{\up,\down}$ are creation and
annihilation operators of the two fermionic species while $b^\dagger$
($b$) creates (annihilates) a closed-channel molecule of mass $M\equiv
m_\up+m_\down$. The atom-molecule interconversion amplitude $g$ is
taken constant up to the momentum cut-off $\Lambda\propto 1/R_e$, and
$\omega_0$ is the bare detuning of the molecule. The quantities $a$
and $R^*$ are related to the parameters of the model
(\ref{eq:hamiltonian}) by \cite{Gurarie2007} (see also Appendix
\ref{app:a})
\begin{equation}\label{Parameters}
a=\frac{\mu g^2}{2\pi}\frac1{\frac{g^2\mu\Lambda}{\pi^2}-\omega_0}, 
\hspace{1cm} R^*=\frac{\pi}{\mu^2g^2}.
\end{equation}

The bare propagators of atoms and closed-channel molecules read
\begin{eqnarray}
G_{\up,\down}(\p,p_0) & = & \frac1{p_0-p^2/2m_{\up,\down}+i0}, \nn \\
D_0(\p,p_0) & = & \frac1{p_0-p^2/2M-\omega_0+i0},
\end{eqnarray}
where $+i0$ slightly shifts the poles of $G$ and $D_0$ into the lower
half of the complex $p_0$-plane. A physical dimer consists of a
closed-channel molecule dressed by open-channel atoms. The
corresponding propagator is given by (see Appendix \ref{app:a})
\begin{eqnarray}
&&\hspace{-0.7cm}D(\p,p_0) \nn \\
&&\hspace{-0.7cm}= \frac{2\pi/\mu}{2\mu R^*\!\left(p_0-\frac{p^2}{2M}+i0 \right)\!+\frac1{a}-\sqrt{2\mu}
\sqrt{-p_0+\frac{p^2}{2M}\!-\!i0}}.
\label{eq:DressedMoleculePropagator}
\end{eqnarray}
The pole of $D(0,p_0)$ determines the dimer binding energy
\begin{equation}
\epsilon_0 = -(\sqrt{1+4R^*/a}-1)^2/8\mu {R^*}^2.
\label{eq:eb}
\end{equation}
Equation (\ref{eq:eb}) interpolates between the two limits: for small
detuning we have $\epsilon_0\simeq -1/2\mu a^2$ and in the regime of
intermediate detuning $\epsilon_0\simeq-1/2\mu R^* a$.

\section{Atom-dimer scattering \label{sec:3body}}

Knowledge of atom-dimer interaction parameters is necessary for the
correct description of an atom-molecule mixture on the BEC side of the
Feshbach resonance. The momentum-space formalism for the three-body
problem with short-range interactions was first demonstrated in the
calculation of the neutron-deuteron scattering length (total spin
$S=3/2$ in this case corresponds to our $\up$-$\up\down$ scattering
problem) \cite{Skorniakov1956}. The coordinate formulation can be
found in Ref.~\cite{Petrov2003} where the atom-dimer scattering length
was obtained in the mass-imbalanced case. Here we extend these results
to higher partials waves, finite collision energies, and finite
Feshbach resonance width.

\begin{figure}[tb]
\includegraphics[width=\hsize,clip]{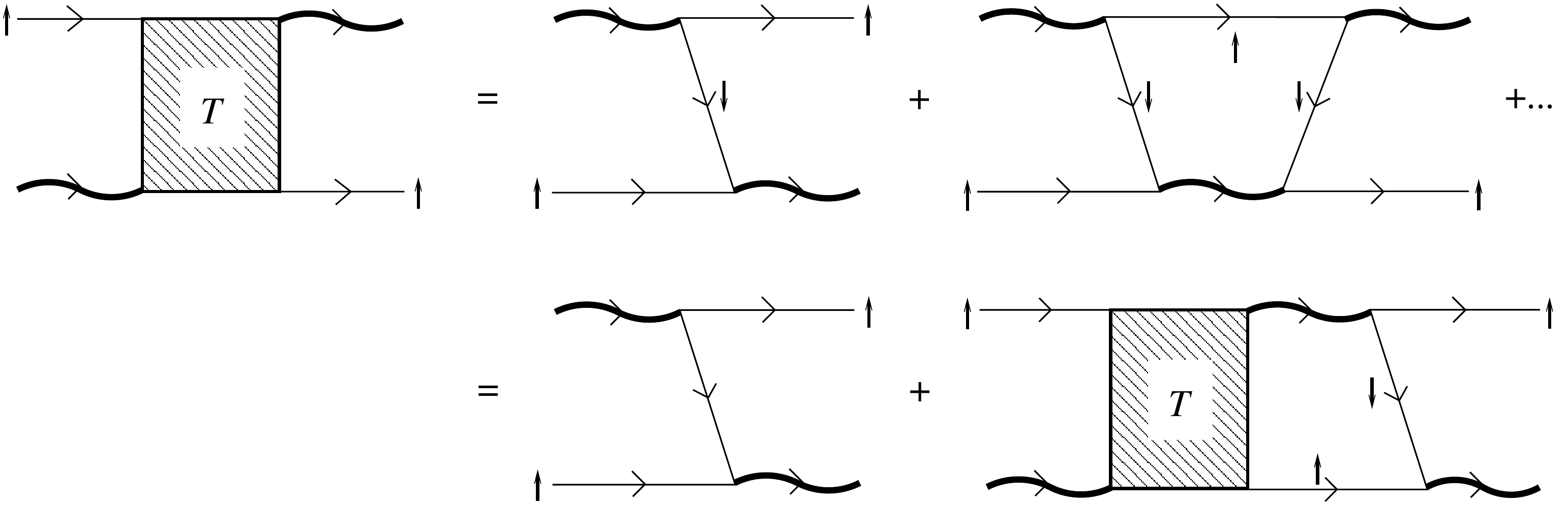}
\caption{Diagrammatic series contributing to the atom-dimer $T$-matrix and a schematic representation of the Skorniakov-Ter-Martirosian integral equation (\ref{eq:t}). External propagators are included to guide the eye, they do not form part of the $T$-matrix. Straight and wavy lines denote atomic and dimer propagators, respectively.}
\label{fig:3body}
\end{figure}

Let us denote the atom-dimer scattering $T$-matrix by $T(\k,k_0;\p,p_0)$, the
arguments of which imply that the incoming four-momenta of the atom
and the molecule are $(\k,k_0)$ and $(-\k,E-k_0)$, and the outgoing
ones are $(\p,p_0)$ and $(-\p,E-p_0)$, respectively. In
Fig.~\ref{fig:3body} we show the diagrammatic series for $T$, the
summation of which results in the Skorniakov-Ter-Martirosian integral
equation \cite{Skorniakov1956} (see also Ref. \cite{Bedaque1998})
\begin{eqnarray}
T(\k,k_0;\p,p_0) & = & -g^2Z\,G_\down(-\k-\p,E-k_0-p_0)
\nn \\ &&
\hspace{-20mm}
-i\int \frac{d^4q}{(2\pi)^4}G_\up(\q,q_0)G_\down(-\p-\q,E-p_0-q_0)
\nn \\ &&
\hspace{0mm}
\times D(-\q,E-q_0) T(\k,k_0;\q,q_0).
\label{eq:t}
\end{eqnarray}
Equation (\ref{eq:t}) is formally identical to the equal-mass
wide-resonance one, the difference being hidden in the propagators and
the factor $Z$, which serves for correct normalization of external
propagators (see Appendix \ref{app:a}). The atom-dimer elastic
scattering amplitude is proportional to the on-shell $T$-matrix:
\begin{equation}
f(\k,\k')=-\frac{\mu_3}{2\pi}T(\k,k^2/2m_\up;\k',k^2/2m_\up),
\end{equation}
where $\mu_3\equiv Mm_\up/(M+m_\up)$ is the reduced mass of the
atom-dimer system, and $k=|\k|=|\k'|$. Hereafter $f$, $f_\ell$,
$\delta_\ell$, and $\sigma_\ell$ refer to the atom-dimer scattering
parameters, the two-atom interaction being described by $a$ and $R^*$.

Integration over $q_0$ in \rfs{eq:t} may be carried out by closing the
complex contour in the lower half plane. The integration picks up the
contribution from the simple pole of $G_\up$ at $q_0=q^2/2m_
\up$. The scattering phase shifts are on-shell quantities and we let
$k_0=k^2/2m_\up$, $p_0=p^2/2m_\up$, and the total energy
$E=k^2/2\mu_3+\epsilon_0$. The remaining on-shell condition,
$|\p|=|\k|$, should be implemented at the end of the calculations.

The kernel of the resulting three-dimensional integral equation has a
simple pole at $|\q|=|\k|$ hidden in the dimer propagator. We make it
explicit by defining functions $\tilde f$ and $h$:
\begin{eqnarray}
\frac{\tilde f(\k,\q)}{q^2-k^2-i0} & = &
\frac{D(\q,E-q^2/2m_\up)}{4\pi g^2Z}
T\left(\k,\frac{k^2}{2m_\up};\q,\frac{q^2}{2m_\up}\right), \nn\\
h(k,q)&=& (k^2-q^2)D(\q,E-q^2/2m_\up)/4\pi.
\end{eqnarray}
Both $\tilde f$ and $h$ are not singular at $|\q|=|\k|$, and $\tilde f$ is chosen such that $\tilde f(\k,\k')=f(\k,\k')$, for $|\k|=|\k'|$. Finally, we note that Eq.~(\ref{eq:t}) conserves angular momentum and, therefore, can be written as a set of decoupled equations for each partial wave
\begin{equation}
\tilde f_\ell(k,p) = h(k,p)\left\{g_\ell(k,p)+\frac2\pi\int_0^\infty
\hspace{-1mm} q^2dq
\frac{g_\ell(p,q)\tilde f_\ell(k,q)}{q^2-k^2-i0}\right\},
\label{eq:inteq}
\end{equation}
where we define
\begin{equation}
g_\ell(k,p)=\frac12\int_{-1}^1 dx P_\ell(x)G_\down(\k+\p,E-k^2/2m_\up-p^2/2m_\up), \nonumber
\end{equation}
where $x$ is the cosine of the angle between $\k$ and $\p$. Partial atom-dimer scattering amplitudes are related to solutions of Eq.~(\ref{eq:inteq}) by the equation $f_\ell (k)=\tilde f_\ell(k,k)$, and the corresponding phase shifts $\delta_\ell$ are deduced from Eq.~(\ref{eq:fphase}).

\begin{figure}[tb]
\includegraphics[width=\hsize,clip]{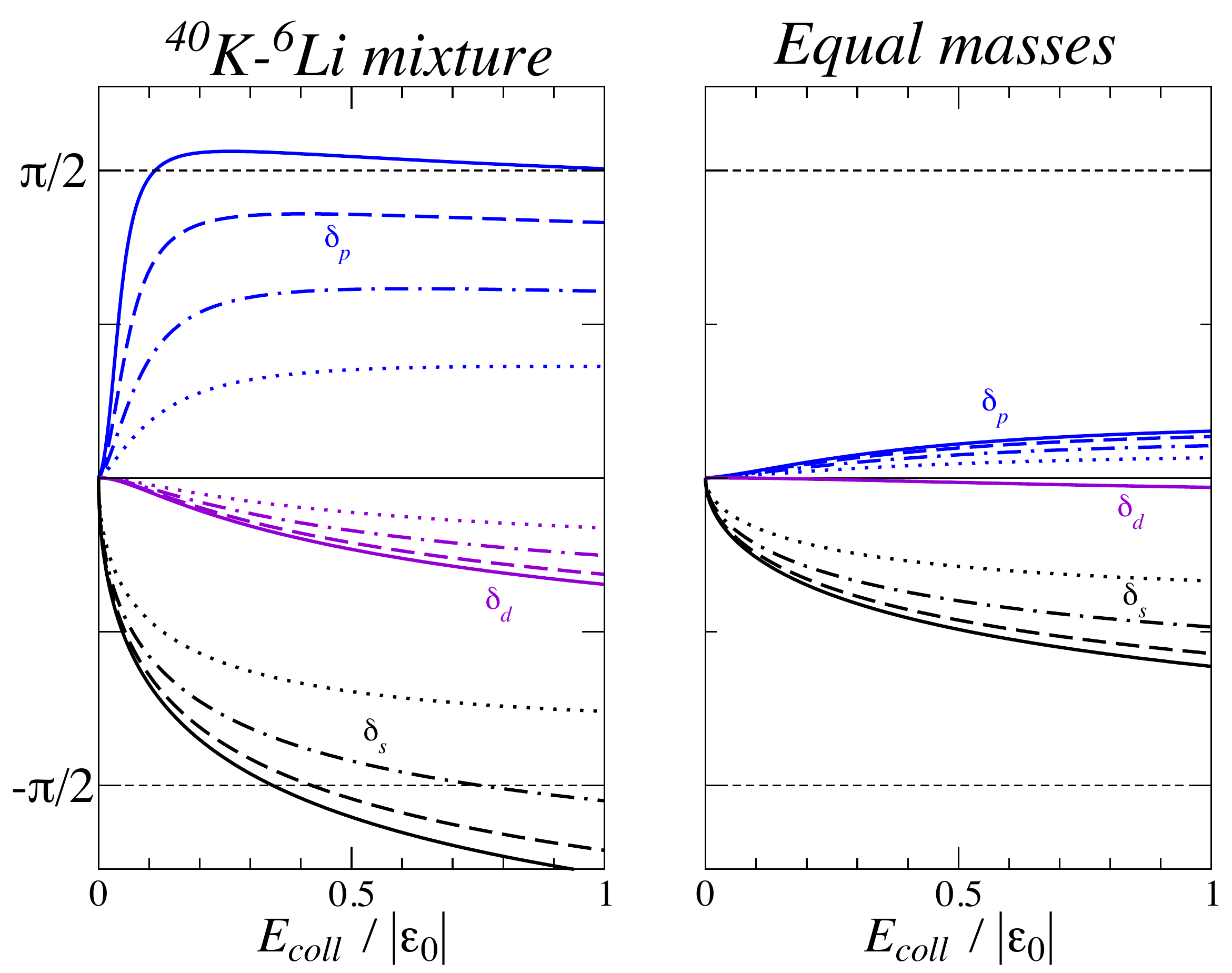}
\caption{(color online). Atom-dimer $s$, $p$, and $d$-wave scattering phase shifts vs. $E_{\mbox{\tiny coll}}/|\epsilon_0|$. Solid, dashed, dot-dashed, and dotted lines correspond to $R^*/a=0$, $1/16$, $1/4$, and $R^*=a$, respectively. In the homonuclear case we show $\delta_d$ only for $R^*=0$.}
\label{fig:phases}
\end{figure}

In Fig.~\ref{fig:phases} we plot the $s$-, $p$-, and $d$-wave phase
shifts as functions of the collision energy $E_{\mbox{\tiny
    coll}}=k^2/2\mu_3$ for different detunings $R^*/a$. We write the
phase shifts as $\delta_0\equiv \delta_s$, $\delta_1\equiv \delta_p$ ,
and $\delta_2\equiv \delta_d$. The results are shown for two mass
ratios: $m_\up/m_\down = 6.64$ (left) and $m_\up/m_\down = 1$
(right). We keep the same vertical scale in both graphs, and one can
see that the atom-dimer interaction in the heteronuclear case is
stronger in every considered channel. Looking at the low-energy
asymptotes of the phase shifts in the wide resonance case ($R^*=0$) we
see that passing from $m_\up/m_\down=1$ to $m_\up/m_\down=6.64$ the
atom-dimer $s$-wave scattering length increases from $a_{\mbox{\tiny
    ad}}\approx1.18 a$ to $a_{\mbox{\tiny ad}}\approx1.98 a$,
consistent with Refs.~\cite{Skorniakov1956} and \cite{Petrov2003}. At
the same time the $p$-wave scattering volume increases by more than an
order of magnitude from $v_{\mbox{\tiny ad}}\approx-0.95a^3$ to
$v_{\mbox{\tiny ad}}\approx-10.1a^3$, which is apparently due to the
vicinity of the resonance at the critical mass ratio
$m_\up/m_\down\approx 8.2$ \cite{Kartavtsev2007}. Although the mass
ratio for the K-Li case is quite a bit smaller, our results indicate
that for sufficiently small detuning one has a strongly marked
$p$-wave K-KLi scattering resonance. Indeed, for $R^*=0$ the $p$-wave
phase shift reaches the unitarity value $\pi/2$ at a relatively small
collision energy $E_{\mbox{\tiny coll}}\approx 0.1 |\epsilon_0|$.

In Fig.~\ref{fig:phases} we also see that the atom-dimer interaction
decreases with detuning. We attribute this to the fact that at larger
$R^*/a$ the light atom spends more time in the closed-channel
molecular state, and consequently contributes less to the atom-dimer
exchange interaction. In a sense, increasing $R^*/a$ is similar to
increasing the mass of the light atom (decreasing the mass ratio): the
heavier the atom, the weaker the exchange interaction.

\begin{figure}[tb]
\includegraphics[width=\hsize,clip]{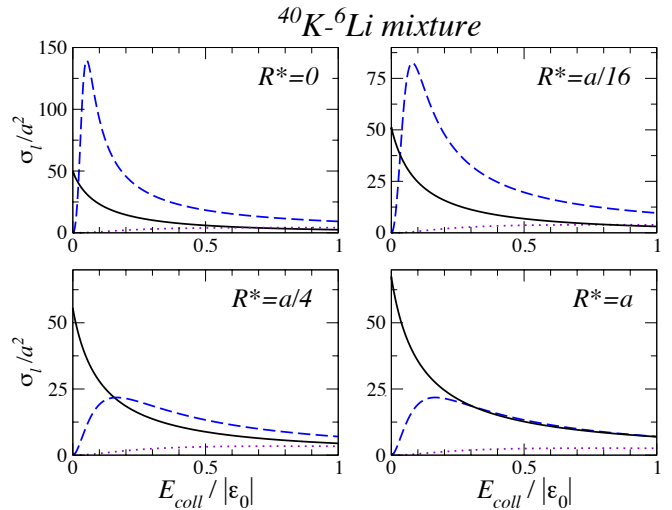}
\caption{(color online). Partial K-KLi atom-dimer cross-sections $\sigma_s$ (solid), $\sigma_p$ (dashed), and $\sigma_d$ (dotted) in units of $a^2$ vs. collision energy for different detunings $R^*/a$.}
\label{fig:cross}
\end{figure}

Although the $p$-wave resonance becomes less pronounced near a narrow
resonance, in a K-Li mixture the $p$-wave atom-dimer interaction can
be strong, which is demonstrated in Fig.~\ref{fig:cross}, where we
plot the partial wave cross-sections
$\sigma_\ell(k)=4\pi(2\ell+1)k^{-2}\sin^2\delta_\ell(k)$. We clearly
see that for detunings $R^*/a\lesssim 1$ the $p$-wave partial
cross-section either exceeds or is comparable to $\sigma_s$ in a wide
range of collision energies.

\begin{figure}[tb]
\includegraphics[width=.85\hsize,clip]{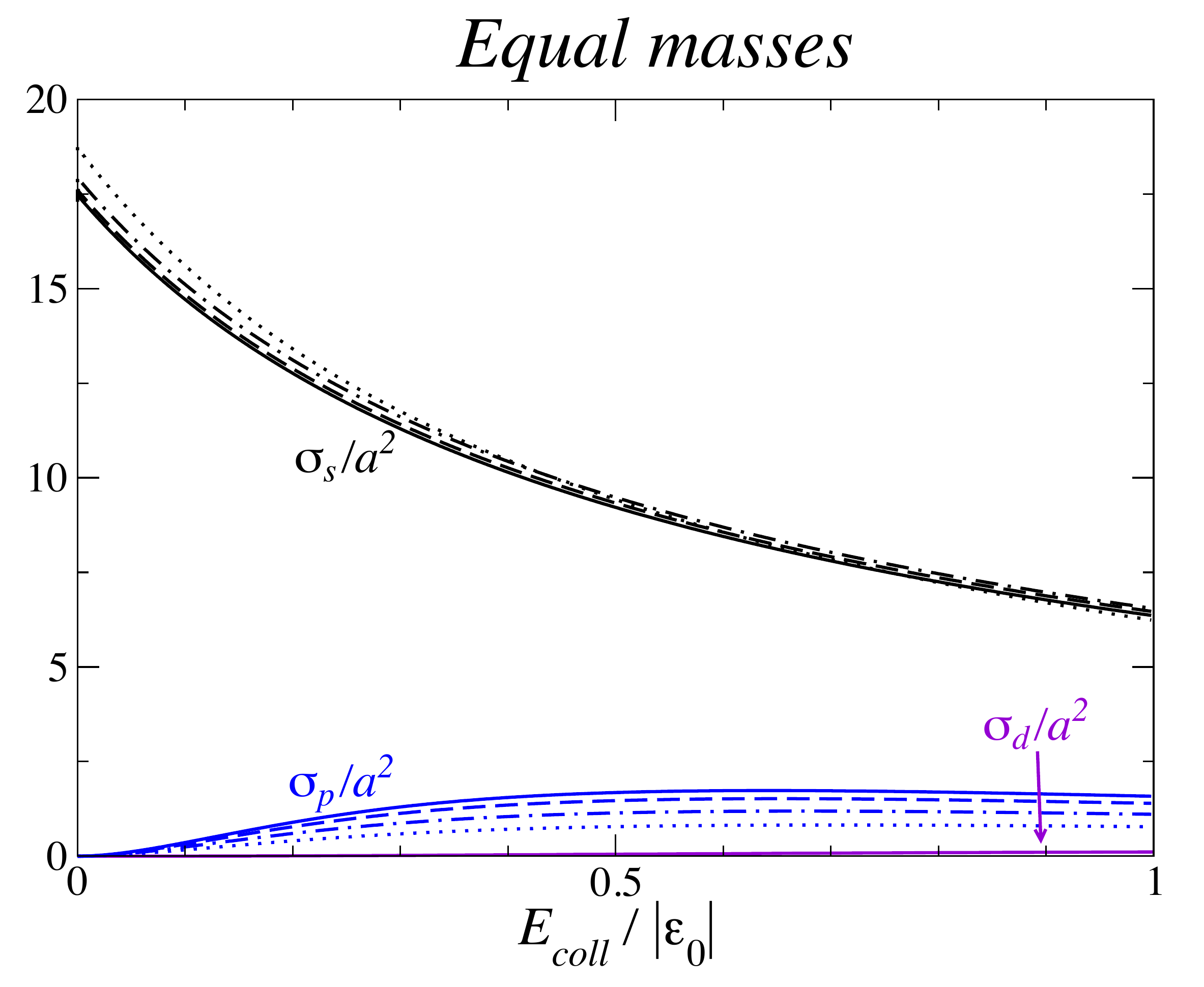}
\caption{(color online). Atom-dimer $s$-wave (black), $p$-wave (blue), and $d$-wave (purple) scattering cross sections for the homonuclear gas. Solid, dashed, dot-dashed, and dotted lines correspond to $R^*/a=0, 1/16, 1/4$, and $R^*=a$, respectively.}
\label{fig:cross2}
\end{figure}

For comparison, in Fig.~\ref{fig:cross2} we present partial atom-dimer
cross-sections in the homonuclear case. We see that the $s$-wave
contribution is always dominant and the functional form of
$\sigma_s/a^2$ is fairly insensitive to the detuning.

We have calculated the scattering parameters for several higher
partial waves. Their contributions rapidly decrease with $\ell$ and it
is worth plotting only the $d$-wave phase shifts (see
Fig.~\ref{fig:phases}) and scattering cross-sections
(Figs.~\ref{fig:cross} and \ref{fig:cross2}). The $d$-wave
contribution is comparable to the $s$- and $p$-wave ones only in the
heteronuclear case and for relatively high collision energies $\sim
|\epsilon_0|$. In Figs.~\ref{fig:phases} and \ref{fig:cross2} the
$d$-wave contribution for the homonuclear case is plotted only for
$R^*=0$ as for finite detunings the curves are even closer to the
horizontal axis.

\begin{figure}[t]
\includegraphics[width=.48\textwidth]{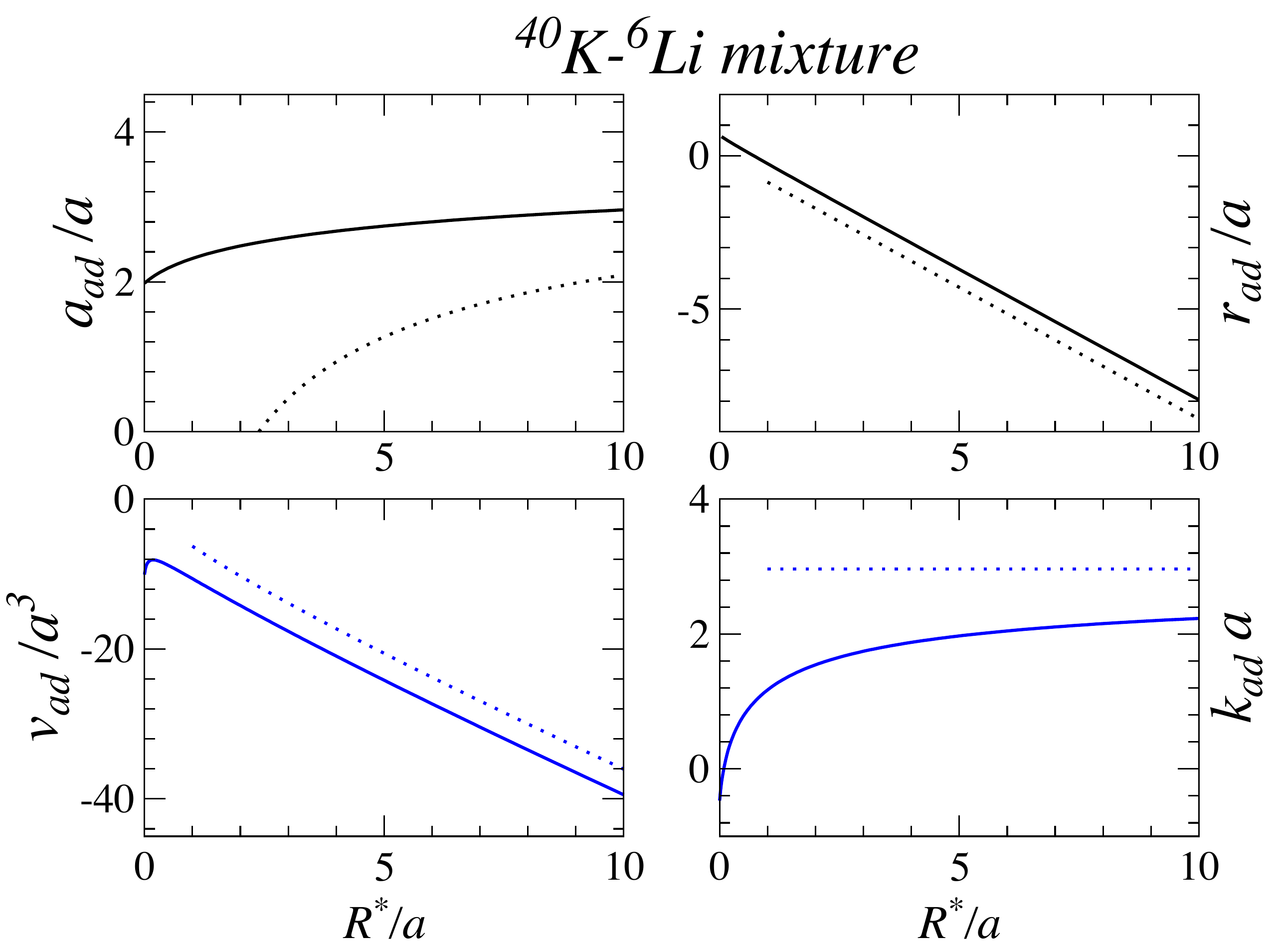}
\caption{(color online). Atom-dimer scattering length $a_{\mbox{\tiny ad}}$, $s$-wave effective range $r_{\mbox{\tiny ad}}$, $p$-wave scattering volume $v_{\mbox{\tiny ad}}$, and the $p$-wave effective range parameter $k_{\mbox{\tiny ad}}$ in units of corresponding powers of $a$ vs. $R^*/a$ for $m_\up/m_\down=6.64$. Solid lines are exact and dotted lines are approximate results (\ref{eq:aad})-(\ref{eq:kad}) valid in the limit $R^*\gg a$.}
\label{fig:a3r3v3k3}
\end{figure}

\begin{figure}[t]
\includegraphics[width=.48\textwidth]{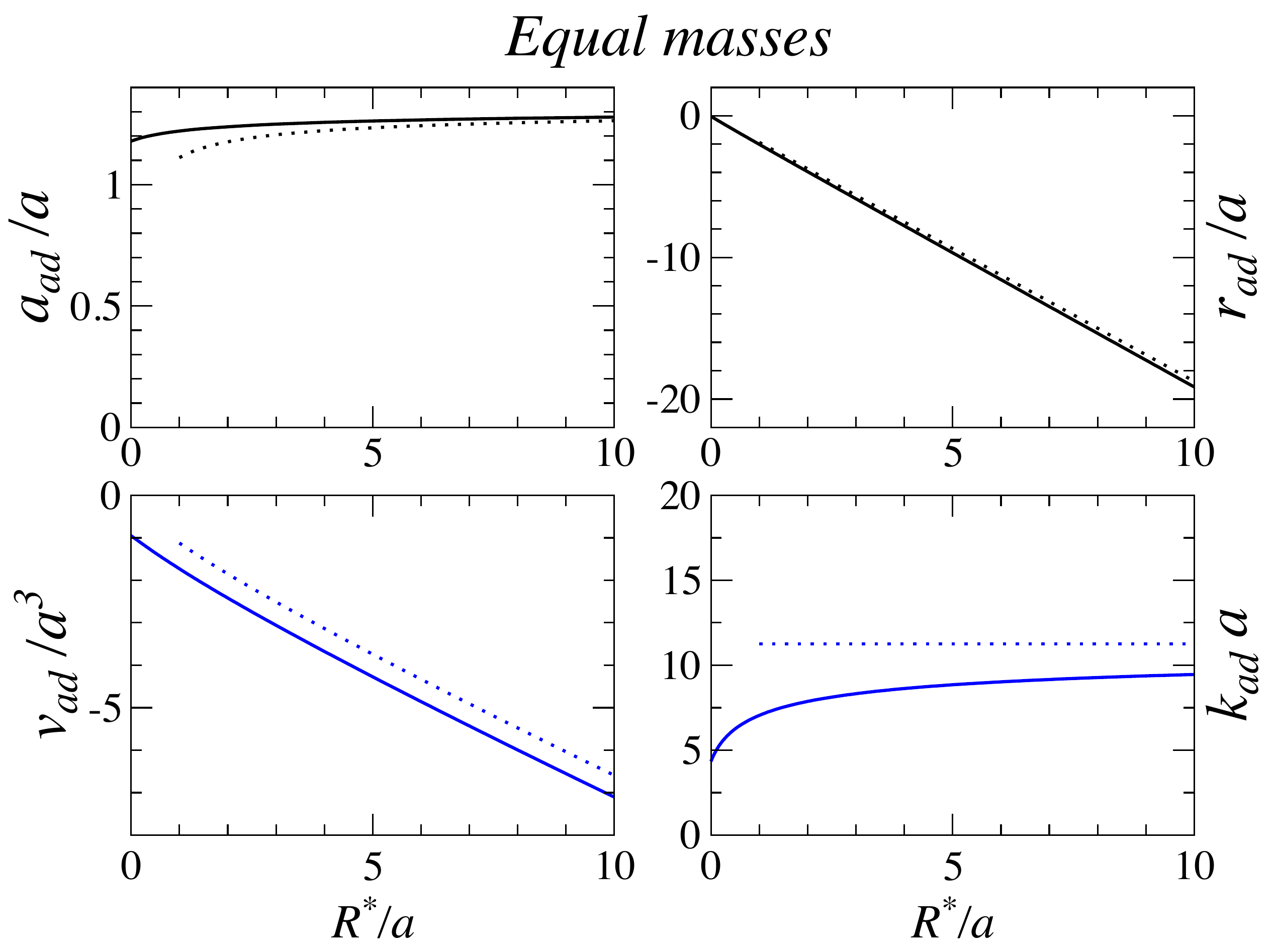}
\caption{(color online). Same as in Fig. \ref{fig:a3r3v3k3} but for the homonuclear case.}
\label{fig:a3r3v3k32}
\end{figure}

We have already discussed the atom-dimer scattering length and
scattering volume for $m_\up/m_\down\approx 6.64$ and $m_\up/m_\down =
1$ in the case $R^*=0$. In Figs.~\ref{fig:a3r3v3k3} and
\ref{fig:a3r3v3k32} we plot these quantities and the effective range
parameters $r_{\mbox{\tiny ad}}$ and $k_{\mbox{\tiny ad}}$ [atom-dimer
analogues of $r_0$ and $k_0$ defined in Eqs.~(\ref{eq:delta0lowk}) and
(\ref{eq:delta1lowk})] versus the detuning $R^*/a$. Dotted lines in
these graphs are obtained by using a perturbation theory in the limit
of a very narrow resonance, $g\rightarrow 0$, when the atom-dimer
$T$-matrix can be obtained by summing the first few diagrams in
Fig.~\ref{fig:3body}. This gives an expansion in powers of
$\sqrt{a/R^*}\ll 1$. The first two terms in the expansion of
$a_{\mbox{\tiny ad}}$ and $v_{\mbox{\tiny ad}}$ and the leading terms
for the effective range parameters $r_{\mbox{\tiny ad}}$ and
$k_{\mbox{\tiny ad}}$ read
\begin{eqnarray}
a_{\mbox{\tiny ad}} & \approx & a\frac{\mu_3}{\mu}\left[1+\frac12\left(1-\frac{\mu_3}{\mu}\right)
\sqrt{\frac{a}{R^*}}\right], \label{eq:aad}\\
r_{\mbox{\tiny ad}} & \approx & -4R^*\frac{\mu}{\mu_3}\left(1-\frac\mu{2\mu_3}\right),\\
v_{\mbox{\tiny ad}} & \approx & -\frac23a^2R^*\frac{\mu_3}{m_\down}
\left[1+\frac32\left(1+\frac{\mu_3}{36m_\down}\right)\sqrt{\frac{a}{R^*}}\right], \\
k_{\mbox{\tiny ad}} & \approx & \frac{12}a\frac{m_\down}{\mu_3}\left(1-\frac\mu{2\mu_3}\right).
\label{eq:kad}
\end{eqnarray}

\subsection{Born-Oppenheimer approximation \label{sec:BO}}

It is instructive to consider the enhancement of the atom-dimer
scattering and the appearance of trimers for sufficiently large mass
ratios in the Born-Oppenheimer approximation \cite{Born1927}. This
method was introduced in Ref.~\cite{Fonseca1979} to study Efimov
physics in the system of one light and two heavy particles. Although
the Born-Oppenheimer approximation is not exact, it serves well to
illustrate the essential physics leading to the resonant enhancement
of the $p$-wave scattering. Here we extend it to the case of a
resonance of finite width.

The method takes advantage of the large mass ratio by assuming that
the state of the light atom adiabatically adjusts itself to the
distance $\R$ between the heavy fermions. The wavefunction of the
light atom can be written in the form
\begin{equation}\label{eq:BOwavefunction}
\psi_{\R,\pm}(\r) \propto \frac{e^{-\kappa_\pm(R)|\r-\R/2|/R}}{|\r-\R/2|}
\pm \frac{e^{-\kappa_\pm(R)|\r+\R/2|/R}}{|\r+\R/2|}.
\end{equation}
It satisfies the free-particle Schr\"odinger equation with the energy 
\begin{equation}\label{eq:BOenergy}
\epsilon_\pm(R)=-\kappa_\pm^2(R)/2m_\down.
\end{equation}
The singularities of $\psi_{\R,\pm}(\r)$ at vanishing $\tilde \r = \r
\pm \R/2$ satisfy the Bethe-Peierls boundary condition \cite{Bethe1935}
\begin{eqnarray}\label{eq:BP}
[\tilde r\psi]'_{\tilde r}/\tilde r\psi\left|\right. {}_{\tilde \r \to 0}&=&i \kappa_\pm(R)\cot\delta_0\left[i\kappa_\pm(R)\right]\nonumber \\
&=& -1/a+R^*\kappa_\pm^2(R),
\end{eqnarray}
where the light-heavy $s$-wave phase shift, again denoted by
$\delta_0$, is calculated at the light-heavy collision energy
$\epsilon_\pm(R)$. The equation for $\kappa_\pm(R)$ is obtained by
applying the boundary condition (\ref{eq:BP}) to the wavefunction
(\ref{eq:BOwavefunction}):
\begin{equation}\label{eq:kappa}
\kappa_\pm(R)\mp \exp\left[-\kappa_\pm(R)R\right]/R= 1/a-R^*\kappa_\pm^2(R).
\end{equation}

The second step of the Born-Oppenheimer method consists of solving the
Schr\"odinger equation for the heavy fermions by using
$\epsilon_\pm(R)$ as the potential energy surface. Let us denote the
corresponding heavy-fermion wavefunction by $\phi(\R)$. Since the
total three-body wavefunction, proportional to the product
$\phi(\R)\psi_{\R,\pm}(\r)$, should be antisymmetric with respect to
the permutation of the heavy fermions, the symmetry of $\phi$ depends
on the choice of sign in Eq.~(\ref{eq:BOwavefunction}). As
$\psi_{\R,+}(\r)$ is symmetric with respect to the permutation
$\R\leftrightarrow -\R$, the heavy-atom wavefunction $\phi$ is
antisymmetric and describes odd atom-dimer scattering
channels. Accordingly, the lower sign in
Eqs.~(\ref{eq:BOwavefunction}-\ref{eq:kappa}) corresponds to even
channels. We see how the composite nature of the dimer leads to the
$\ell$-dependent effective atom-dimer potentials: by solving
Eq.~(\ref{eq:kappa}) one arrives at a purely attractive
$\epsilon_+(R)$ for odd channels and purely repulsive $\epsilon_-(R)$
for even ones.

From the viewpoint of the radial Schr\"odinger equation it is
convenient to introduce the total effective potential for each
$\phi_\ell (R)$:
\begin{equation}
V_\ell(R)=\epsilon_{(-1)^{\ell+1}}(R)-\epsilon(\infty)+\ell(\ell+1)/m_\up R^2,
\end{equation}
which includes the centrifugal barrier and shifts the threshold to
zero by subtracting the dimer binding energy
\begin{equation}\label{eq:BObindinenergy}
\epsilon(\infty)=-(\sqrt{1+4R^*/a}-1)^2/8m_\down R^{*2}.
\end{equation}
In the limit $m_\up\gg m_\down$ Eq.~(\ref{eq:BObindinenergy}) reduces
to Eq.~(\ref{eq:eb}).

\begin{figure}
\includegraphics[width=.45\textwidth]{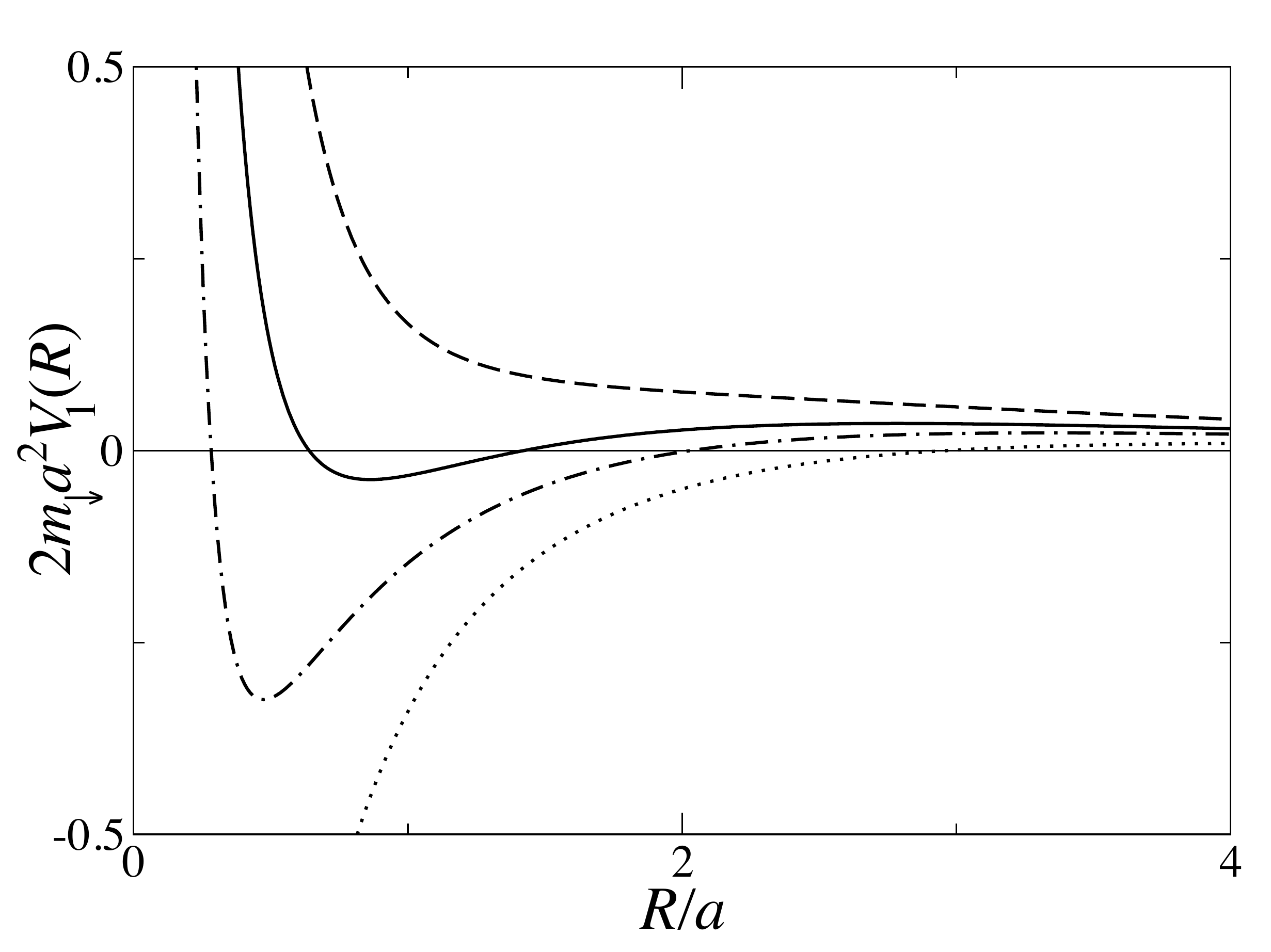}
\caption{The Born-Oppenheimer atom-dimer effective potential $V_1(R)$ [in units of $1/2m_\down a^2$] in the wide resonance case ($R^*=0$) for mass ratios $m_\up/m_\down=5$ (dashed), $m_{\mbox{K}}/m_{\mbox{Li}}$ (solid), 8.2 (dash-dotted), and 13.6 (dotted).}
\label{fig:pot}
\end{figure}

For the $p$-wave atom-dimer interaction the central issue is the
competition between the attractive exchange potential
$\epsilon_+\propto 1/m_\down$ and the repulsive centrifugal barrier,
which is inversely proportional to $m_\up$. In Fig.~\ref{fig:pot} we
show $V_1(R)$ in the limit of vanishing detuning for different mass
ratios. Remarkably, for $m_\up/m_\down\sim m_{\mbox{\tiny K}}
/m_{\mbox{\tiny Li}}$ this potential, being repulsive in both limits
$R\ll a$ and $R\gg a$, develops a well at distances $R\sim a$. For
$m_\up/m_\down > 8.2$ the depth of this well is enough to accomodate a
trimer state with unit angular momentum \cite{Kartavtsev2007} and for
somewhat smaller mass ratios the presence of the well leads to the
resonant enhancement of the $p$-wave interaction.

\begin{figure}
\includegraphics[width=.45\textwidth]{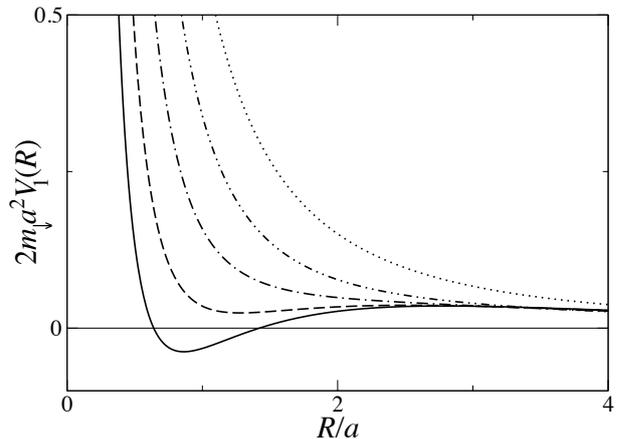}
\caption{The Born-Oppenheimer atom-dimer effective potential $V_1(R)$ [in units of $1/2m_\down a^2$] in the K-Li case for $R^*=0$ (solid), $R^*=a/16$ (dashed), $R^*=a/4$ (dash-dotted), and $R^*=a$ (double-dot dashed). The dotted line is the centrifugal barrier.}
\label{fig:pot2}
\end{figure}

The effect of finite $R^*$ is to decrease the strength of the exchange
potentials $\epsilon_\pm$. In Fig.~\ref{fig:pot2} we show $V_1(R)$ in
the K-Li case for different values of the detuning $R^*/a$. One can
see that the $p$-wave attraction becomes less pronounced and the well
eventually disappears with increasing $R^*/a$.

It is important to distinguish the $p$-wave trimer for
$m_\up/m_\down\gtrsim 8.2$ from Efimov trimers. The former exists only
for $a>0$ and is a result of the peculiar competition between the
exchange potential and the centrifugal force at distances of the order
of $a$, which determines its size. In contrast, the Efimov effect
occurs at larger mass ratios, $m_\up/m_\down > 13.6$, when the
effective potential at distances $R\ll a$ is no longer
repulsive. Then, in the Born-Oppenheimer description the heavy atoms
fall to the center in an attractive $1/R^2$-potential. This is
accompanied by the appearance of an infinite set of Efimov states,
irrespective of the sign of $a$.

Atom-dimer scattering in even channels is described by the potential
$\epsilon_+(R)$, which is defined at distances $R>a$. It has a form of
a purely repulsive soft-core potential, which increases with the mass
ratio and decreases with $R^*$, consistent with the exact results
above on $s$- and $d$-wave atom-dimer scattering.

\subsection{Interference of $s$- and $p$-waves}

Now we would like to discuss one of the implications of the channel
dependent atom-dimer interaction, a peculiarity, which is strongly
pronounced in the K-Li mixture. In this case the $p$- and $s$-wave
phase shifts are comparable in magnitude and can be large (see
Fig.~\ref{fig:cross}). It is thus not necessary to go to very high
collision energies for observing the quantum interference between
these partial waves \cite{note:high}.

Let us consider a {\em gedankenexperiment} in which a cold thermal
cloud of KLi dimers collides with a cloud of K atoms at collision
energies below the dimer break-up threshold. The measurable quantity
is then the angular distribution of scattered dimers (or atoms), which
is proportional to the differential cross-section. We can write it in
terms of the phase shifts by using Eqs.~(\ref{eq:f}) and
(\ref{eq:fphase}):
\begin{eqnarray}
\frac{d\sigma}{d\Omega} & = & \frac1{k^2}\left[\sin^2\delta_s+
6\cos(\delta_p-\delta_s)\sin\delta_s\sin\delta_p\cos\theta
\right.\nn\\ &&\hspace{6mm}+\left.
9\sin^2\delta_p\cos^2\theta\right]+\dots
\label{eq:diffsigma}
\end{eqnarray}
Here $k$ is the relative atom-dimer momentum and the angle $\theta$ is
measured with respect to the collision axis, which we denote by
$\z$. The dots signify the contribution of higher partial waves. We
have checked that they can be safely ignored.

\begin{figure}[bt]
\includegraphics[width=.49\textwidth]{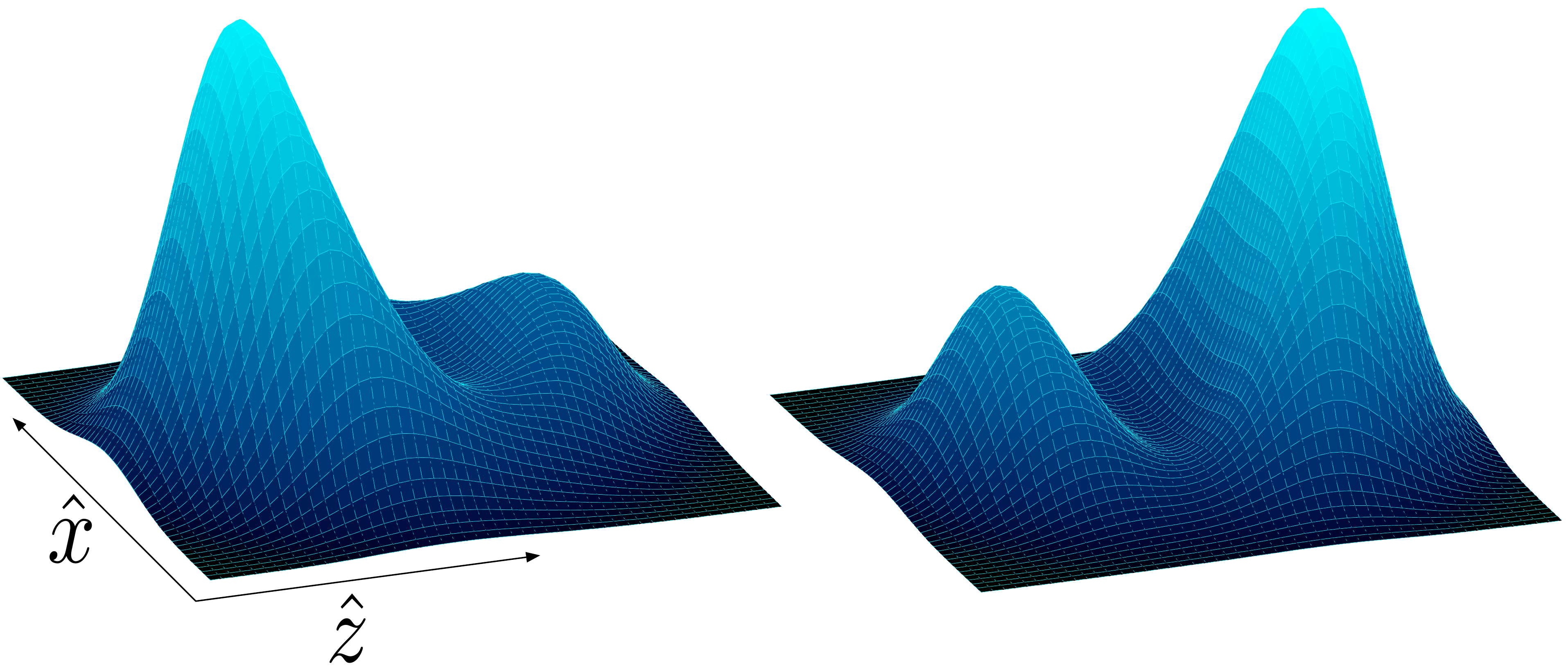}
\caption{(color online). The integrated column density for K-KLi atom-dimer scattering in arbitrary units. The collision energies are $E_{\mbox{\tiny coll}}=0.05|\epsilon_0|$ (left) and $E_{\mbox{\tiny coll}}=0.25|\epsilon_0|$ (right). In both cases $R^*=a/4$. Backward direction corresponds to negative $z$. For presentation purposes we imitate a small thermal smear.}
\label{fig:scat}
\end{figure}

The first term on the right hand side in Eq.~(\ref{eq:diffsigma})
gives the well-known spherically symmetric scattering halo. The last
term corresponds to the pure $p$-wave scattering. It contributes
equally to the forward ($0<\theta<\pi/2$) and backward
($\pi/2<\theta<\pi$) directions, but vanishes in the direction
perpendicular to $\z$. The second (interference) term favors either
backward or forward scattering. In Fig.~\ref{fig:scat} we simulate an
absorption image (column density) of scattered particles that
initially moved in the positive $z$ direction. In the K-Li case
backward scattering dominates at small collision energies while forward
scattering is favored at higher energies, when $\delta_p-\delta_s>\pi/2$.

\begin{figure}
\includegraphics[height=.34\textwidth]{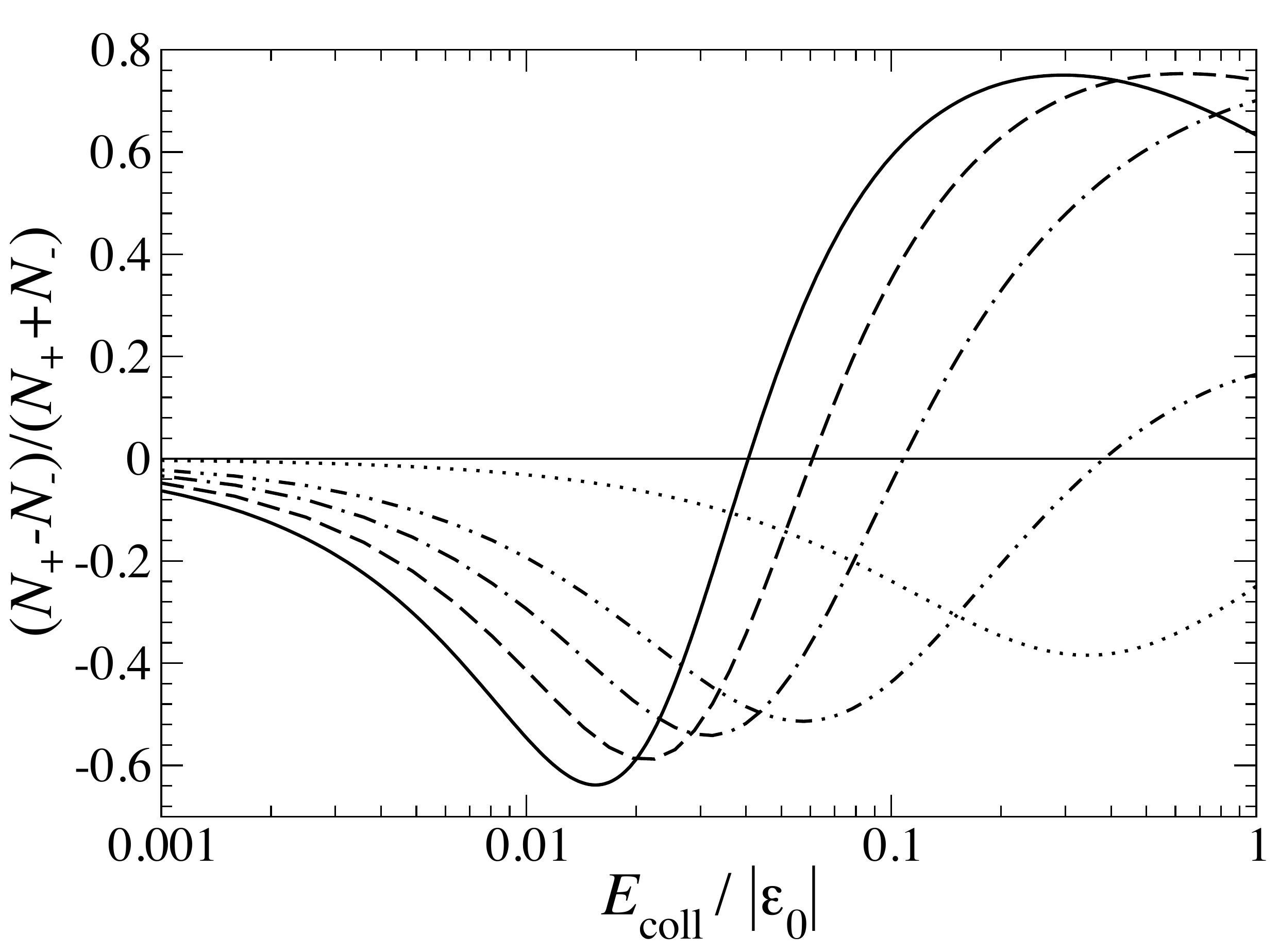}
\caption{The contrast vs. collision energy for detuning $R^*/a$ equal to 0 (solid), 1/16 (dashed), 1/4 (dot-dashed), and 1 (double-dot dashed). Dotted line is the homonuclear case result for $R^*=0$.}
\label{fig:contrast}
\end{figure}

In Fig.~\ref{fig:contrast} we plot the contrast, defined as the
normalized difference between the numbers of particles scattered
forward, $N_+$, and backward, $N_-$, as a function of collision energy
for different detunings $R^*/a$. For comparison we also
present the homonuclear wide-resonance case (dotted line). We see that in this
case backward scattering always dominates, the highest contrast
achieved for $E_{\mbox{\tiny coll}}\approx |\epsilon_0|/3$.

The collision experiment described above requires the ability of
manipulating atoms and molecules individually, which points to an
advantage of heteronuclear mixtures -- in the heteronuclear case
different atomic species feel optical potentials in a different
manner, and this obviously holds for the two components of the
corresponding atom-molecule mixture.

As far as the energy scale is concerned, in the K-Li case with
$R^*=a=100$ nm the dimer binding energy given by Eq.~(\ref{eq:eb})
equals $|\epsilon_0|\approx 1.8$ $\mu$K, and it decreases to 200 nK
when $a = 400$ nm. We also mention, for reference, that the relative
atom-dimer velocities corresponding to the collision energy
$E_{\mbox{\tiny coll}}= |\epsilon_0|$ in these two cases equal 3.7
cm/s and 1.2 cm/s respectively.

\section{Dimer-dimer scattering \label{sec:4body}}

Dimer-dimer interaction parameters are crucial for the description of
the BCS-BEC crossover in the BEC-limit, i.e. when the gas of molecules
is dilute. In the lowest order the chemical potential, condensate
depletion, and speed of sound in the BEC of dimers are determined from
the density and dimer-dimer scattering length $a_{\mbox{\tiny dd}}$ in
the same manner \cite{Levinsen2006} as in the usual Bogoliubov theory
of dilute Bose gases.

In the case of a homonuclear mixture near a wide resonance the
dimer-dimer scattering length equals $a_{\mbox{\tiny dd}}\approx 0.6
a$. This number was obtained in Ref.~\cite{Petrov2004} by solving an
integral equation derived directly from the four-body Schr\"odinger
equation in coordinate space. Later the result was confirmed by
diagrammatic approaches \cite{Levinsen2006,Brodsky2005} and by
Monte-Carlo and variational techniques \cite{Astrakharchik2004,Stecher2007}. In the
heteronuclear case the molecule-molecule scattering length was also
calculated in the case of a wide interspecies resonance
\cite{Petrov2005,Stecher2007,Levinsen2007a}. Von Stecher {\em et al.}
\cite{Stecher2007} also computed the dimer-dimer effective range. The
inelastic scattering and the formation of Efimov trimers in
dimer-dimer collisions for $m_\up/m_\down>13.6$ is discussed in
Ref.~\cite{Marcelis2008}.

A qualitative summary of the results cited above is that the dimer-dimer
interaction can be thought of as a soft-core repulsion. It strengthens
with the mass ratio, but is always weaker than the $s$-wave repulsion
between a dimer and a heavy atom. This picture can be understood from
the Born-Oppenheimer analysis, when one assumes that the wavefunction
of the two light fermions is given by the antisymmetrized product of
$\psi_{\R,+}(\r_1)$ and $\psi_{\R,-}(\r_2)$ [see
Eq.~(\ref{eq:BOwavefunction})], which is antisymmetric under the
permutation of the heavy fermions. Accordingly, the heavy-atom part of
the wavefunction should be symmetric, consistent with the fact that
only even scattering channels are allowed between identical
bosons. The Born-Oppenheimer potential energy surface is given,
independent of the angular momentum, by the sum
$\epsilon_+(R)+\epsilon_-(R)$. It increases with decreasing the mass
of the light atom and is repulsive, but not as strong as the
atom-dimer $s$-wave potential $\epsilon_-(R)$.

As far as we know, higher partial waves in scattering of bosonic
dimers have not been studied, but it has been shown that the ground
state of four $\up-\up-\down-\down$ fermions in an anisotropic
harmonic potential has zero angular momentum, independent of $a$ and
the mass ratio \cite{Stecher2007,BlumePrivate}. Besides, the qualitative
Born-Oppenheimer analysis does not provide arguments for any resonant
enhancement of higher partial waves. We thus conjecture that the
$s$-wave channel should dominate the dimer-dimer interaction, at least
for sufficiently small collision energies.

The aim of this section is to compute the dimer-dimer $s$-wave
scattering length $a_{\mbox{\tiny dd}}$ for the homonuclear and
heteronuclear cases (having in mind the Li-K mixture) taking into
account the finite width of the Feshbach resonance. Our derivation
follows Refs.~\cite{Levinsen2006,Levinsen2007a} where the problem was
studied in the regime of small detuning.

\begin{figure}[t]
\includegraphics[width=.49\textwidth]{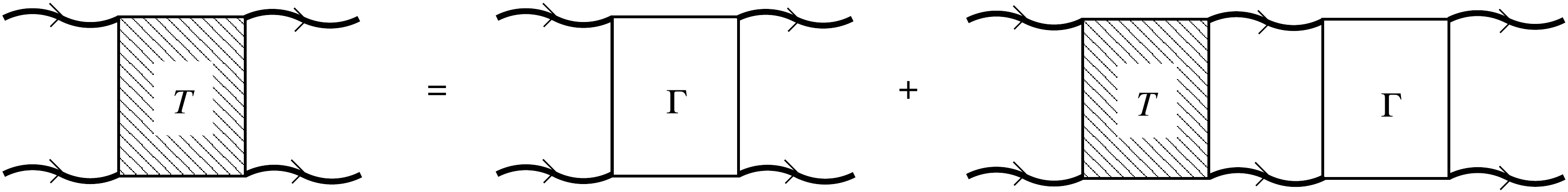}
\caption{Diagrammatic representation of the integral equation (\ref{eq:TGamma}), which relates the dimer-dimer $T$-matrix with the sum of all the two-dimer irreducible diagrams, $\Gamma$.}
\label{fig:4body}
\end{figure}

We consider the scattering of two dimers with four-momenta
$(\0,\epsilon_0)$ into dimers with $(\pm\p,\epsilon_0\pm p_0)$ and
project onto the $s$-wave (average over directions of $\p$). The
four-body $T$-matrix with these kinematics is denoted
$T(p,p_0)$. Similarly to the three-body case the four-body $T$-matrix
consists of an infinite sum of diagrams, which may again be reduced
to integral equations. In order to perform this summation, we first
construct the sum of two-dimer irreducible diagrams beginning and
ending in two dimer propagators. These are the diagrams that cannot be
divided in two by cutting only one pair of dimer propagators (external
lines are excluded from the summation). The corresponding $s$-wave
averaged sum is denoted by $\Gamma(q,q_0;p,p_0)$, where the
four-momenta of the incoming [outgoing] dimers equal $(\pm
\q,\epsilon_0\pm q_0)$ [$(\pm\p,\epsilon_0\pm p_0)$]. The equation for
the $T$-matrix then reads
\begin{eqnarray}
T(p,p_0) & = & g^4Z^2\Gamma(0,0;p,p_0)
+\frac i{4\pi^3}\int q^2dq\,dq_0\,T(q,q_0)\nn \\ && 
\hspace{-10mm}\times \Gamma(q,q_0;p,p_0)
D(q,\epsilon_0+q_0)D(q,\epsilon_0-q_0)
\label{eq:TGamma}
\end{eqnarray}
and is illustrated in Fig.~\ref{fig:4body}. As in the three-body case,
the prefactor of the first term on the right hand side serves for the
correct normalization of external propagators. In order to avoid poles
and branch cuts we solve Eq.~(\ref{eq:TGamma}) by rotating the contour
of the $q_0$-integration to the imaginary axis \cite{Levinsen2006}.

\begin{figure}[t]
\includegraphics[width=.49\textwidth]{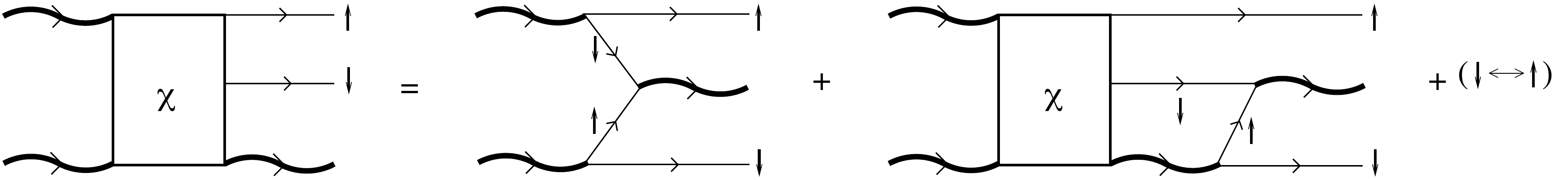}
\caption{The integral equation satisfied by $\chi$, the sum of the two-dimer irreducible diagrams, in which one of the outgoing dimers is split into
  its constituent parts.}
\label{fig:chi}
\end{figure}

\begin{figure}[tb]
\includegraphics[width=.49\textwidth]{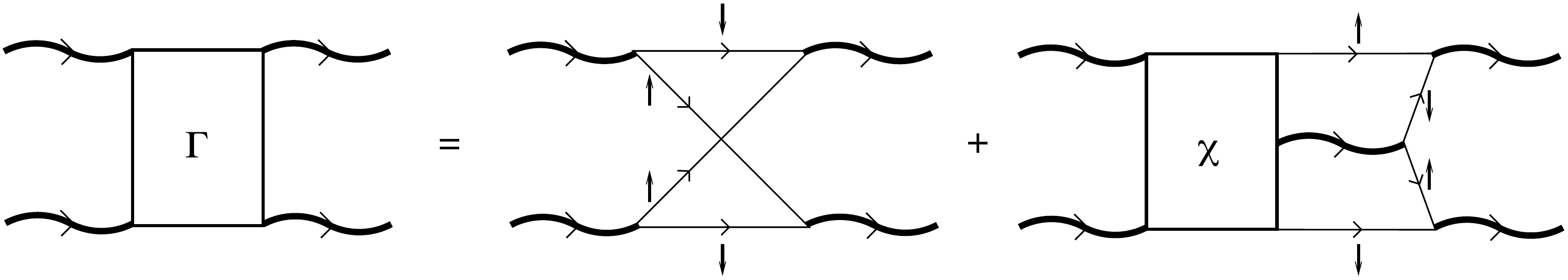}
\caption{$\Gamma$ expressed in terms of $\chi$ (see text).}
\label{fig:gammachi}
\end{figure}

The sum of the two-dimer irreducible diagrams is calculated as
follows. We first sum the two-dimer irreducible diagrams which end in
a dimer and two fermionic atoms. We denote the sum of such two-dimer
irreducible diagrams by $\chi(q,q_0;\p_1,\p_2)$, where the two
incoming dimers have four-momenta $(\pm \q,\epsilon_0\pm q_0)$, and
the outgoing $\up$ [$\down$] atom may be put on-shell with
four-momentum $(\p_1,p_1^2/2m_\up)$ $[(\p_2,p_2^2/2m_\down)]$. By
energy-momentum conservation the outgoing dimer then has four-momentum
$(-\p_1-\p_2,2\epsilon_0-p_1^2/2m_\up-p_2^2/2m_\down)$. In
Fig.~\ref{fig:chi} we illustrate the integral equation satisfied by
$\chi$. The equation itself reads
\begin{eqnarray}
\chi(q,q_0;\p_1,\p_2) & = & -\int\frac{d\Omega_{\vec q}}{4\pi}\left\{
G_\down\left(\q-\p_1,\epsilon_0+q_0-\frac{p_1^2}{2m_\up}\right)
\right. \nn \\ && \hspace{-24mm} \left. \times
G_\up\left(-\q-\p_2,\epsilon_0-q_0-\frac{p_2^2}{2m_\down}\right)+\left[(\q,q_0)
\leftrightarrow -(\q,q_0)\right]\right\}
\nn \\ &&  \hspace{-27mm}
-\int\hspace{-1mm}\frac{d^3Q}{(2\pi)^3}\hspace{-1mm}\left\{
  G_\up\left(\Q+\p_1+\p_2,2\epsilon_0-\frac{Q^2}{2m_\down}-
    \frac{p_1^2}{2m_\up}-\frac{p_2^2}{2m_\down}\right)
\right. \nn \\ &&  \hspace{-18mm} \left.\times
 D\left(\Q+\p_1,2\epsilon_0-\frac{Q^2}{2m_\down}-\frac{p_1^2}{2m_\up}\right)
  \chi(q,q_0;\p_1,\Q)
\right.\nn \\ && \left. \hspace{-14mm}
+G_\down\left(\Q+\p_1+\p_2,2\epsilon_0-\frac{Q^2}{2m_\up}
-\frac{p_1^2}{2m_\up}-\frac{p_2^2}{2m_\down}\right)
\right. \nn \\ &&  \hspace{-18mm} \left.\times
D\left(\Q+\p_2
,2\epsilon_0-\frac{Q^2}{2m_\up}-\frac{p_2^2}{2m_\down}\right)
\chi(q,q_0;\Q,\p_2)
\right\},\nn \\ &&
\label{eq:onshellchi}
\end{eqnarray}
where the frequency integration in the closed loop of the iterated
term is already performed. The configurational space of
Eq.~(\ref{eq:onshellchi}) is in fact three-dimensional. It consists of
the moduli of the vectors $\p_1$ and $\p_2$, and the angle between
them. The pair $(q,q_0)$ enters parametrically. In order to express
$\Gamma$ in terms of $\chi$, it is advantageous to separate out the
simplest diagram, in which the dimers exchange identical atoms. Then,
the remaining diagrams in $\Gamma$ are obtained by closing the
fermionic loop in $\chi$ (see Fig. \ref{fig:gammachi}).

The relation between $\Gamma$ and $\chi$ is
\begin{eqnarray}
  \Gamma(q,q_0;p,p_0) & = & \Gamma^{(0)}(q,q_0;p,p_0)
  \nn \\ &&
\hspace{-24mm}
  -\frac12\int\frac{d^3p_1}{(2\pi)^3} \frac{d^3p_2}{(2\pi)^3}
  \left\{G_\down\left(\p-\p_1,\epsilon_0+p_0-\frac{p_1^2}{2m_\up}\right)
\right. \nn \\ && \hspace{-22mm}\left.\times
    G_\up\left(\p+\p_2,\epsilon_0-p_0-\frac{p_2^2}{2m_\down}\right)
    +\left[(\p,p_0)\leftrightarrow -(\p,p_0)\right]\right\}
\nn \\ && \hspace{-21mm}  \times
  D\left(\p_1+\p_2,2\epsilon_0-\frac{p_1^2}{2m_\up}-\frac{p_2^2}{2m_\down}\right)
  \chi(q,q_0;\p_1,\p_2),
\end{eqnarray}
where the factor $\frac12$ is needed for correct counting of
diagrams. The quantity $\Gamma^{(0)}$ is the first diagram on the
right hand side of Fig.~\ref{fig:gammachi} and is given in Appendix
\ref{app:b}.

\begin{figure}[tb]
\includegraphics[width=.49\textwidth]{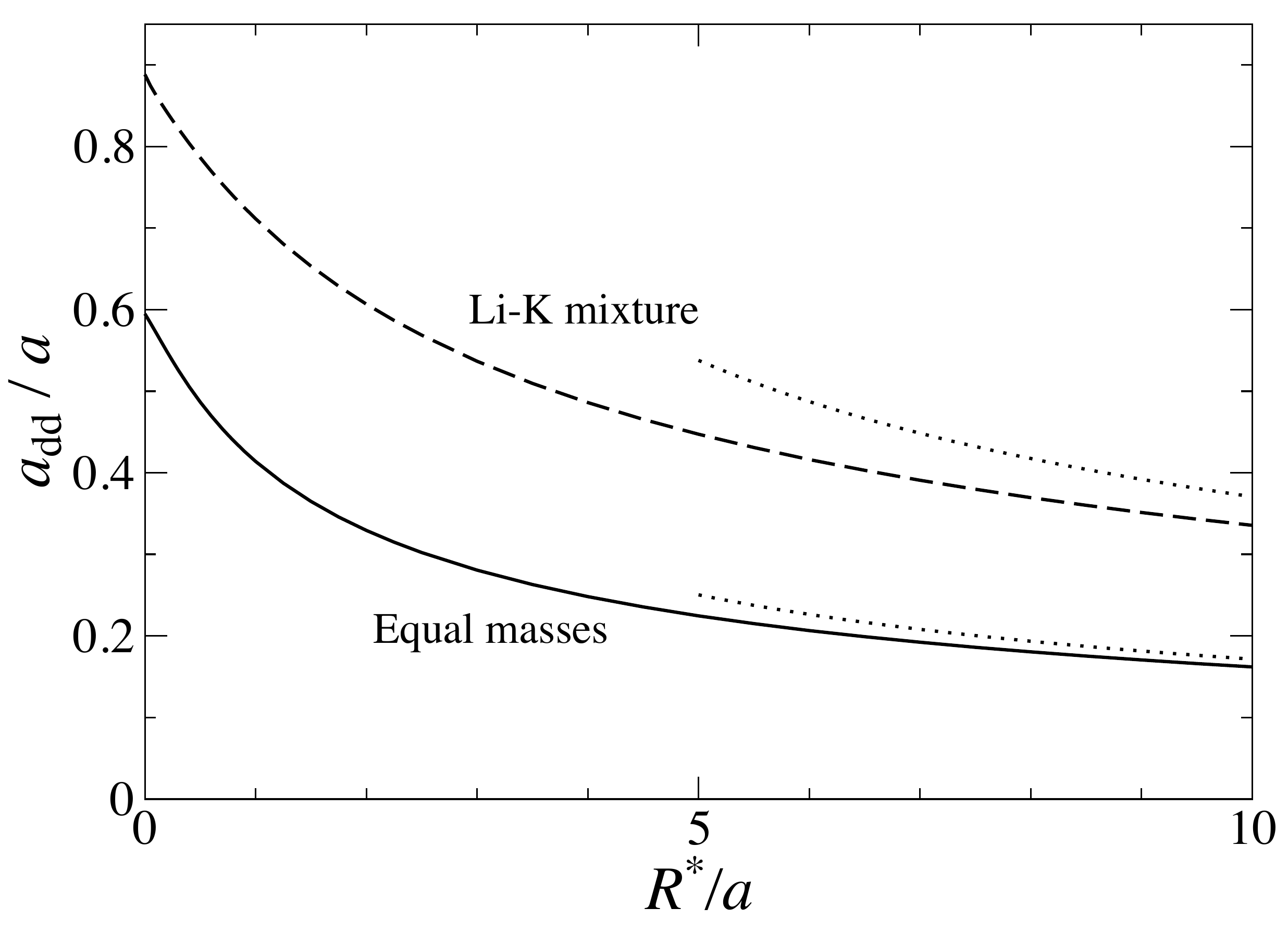}
\caption{The dimer-dimer scattering length vs. $R^*/a$ for equal
  masses (solid line) and for the Li-K mixture (dashed). The dotted
  lines correspond to the asymptote (\ref{eq:addlimit}) valid in the
  limit $R^*\gg a$.}
\label{fig:a4}
\end{figure}

The dimer-dimer scattering length is related to the $T$-matrix by
\begin{equation}
a_{\mbox{\tiny dd}} = \frac M{4\pi}T(0,0).
\end{equation}
Fig.~\ref{fig:a4} shows our results for the dimer-dimer scattering
length in the equal mass case and for the Li-K mixture. In the limit
of small detuning we recover the results
\cite{Petrov2004,Petrov2005,Stecher2007,Levinsen2007a}
\begin{eqnarray}
&a_{\mbox{\tiny dd}}=0.60,&\hspace{1cm}m_\up/m_\down=1, \\
&a_{\mbox{\tiny dd}}=0.89,&\hspace{1cm}m_\up/m_\down=m_{\mbox{\tiny
    K}}/m_{\mbox{\tiny Li}}.
\end{eqnarray}
In the opposite limit the diagrammatic expansion becomes perturbative
as is the case for the atom-dimer problem discussed in
Sec.~\ref{sec:3body}. The dominant contribution to the dimer-dimer
$T$-matrix is provided by $\Gamma^{(0)}$. Including also the next
order, we find \cite{note:addlargerstar}
\begin{equation}
\frac{a_{\mbox{\tiny dd}}}a =
\frac{M}{8\mu}\sqrt{\frac{a}{R^*}}+\frac{a}{R^*}\times
\left\{\begin{array}{lc}0.13, & m_\up=m_\down\\0.23,&
m_\up/m_\down=6.64 
\end{array}
\right.,\,R^*\gg a.
\label{eq:addlimit}
\end{equation}
The first term on the right hand side of Eq.~(\ref{eq:addlimit}) has
been derived in the equal-mass case in Ref.~\cite{Gurarie2007}.

\section{Relaxation rates \label{sec:relaxation}}

The weakly bound dimers that we are considering are in fact molecules
in the highest rovibrational state. They can undergo relaxation into
deep bound states in collisions with each other or with unbound
atoms. The process is local as it requires at least three atoms to
approach each other to a distance comparable to the size of the future
molecular state, i.e. $\sim R_e$. The released binding energy is of
the order of $1/m_\down R_e^2$ and is much larger than all other energy
scales in the problem including the height of the trapping
potential. Thus, the relaxation products are lost.

Although the relaxation is a short-range phenomenon, it can be treated
in the zero-range approximation. For wide resonances in the Efimov
case, i.e. for bosons or for fermions with $m_\up / m_\down >13.6$, the Efimov physics is well
described by the motion of three atoms in an effective attractive
$1/R^2$ potential \cite{Efimov1970,NielsenPhysRep2001}. The three-body wavefunction can be separated in an
incoming wave and an outgoing one, and the relaxation process is taken into account by adding an imaginary
part to the three-body parameter \cite{braaten07}. It fixes the ratio of the corresponding incoming and
outgoing fluxes. The physical range of the potential $R_e$ does not
enter the resulting relaxation rate constant.

The suppression of relaxation in the non-Efimovian cases ({\em i.e.} the
$\up\up\down$ system of fermions with $m_\up / m_\down <13.6$)
originates from the centrifugal barrier for identical fermions, which,
in turn, leads to the repulsive effective three-body $1/R^2$
potential \cite{Efimov1973,IncaoEsry2005}. In order to recombine, the atoms have to tunnel under this
barrier to distances $\sim R_e$. The zero-range approach in this case
is perturbative. It uses the unperturbed few-body wavefunction to
predict the probability of finding three atoms at small distances and
gives the functional dependence of the relaxation rate constant on the
scattering length for a given mass ratio
\cite{Petrov2005a,Petrov2005}. If the relaxation rate constant is
known for a certain $a$, one can predict its value for any other $a
\gg R_e$.

To be more specific, let us demonstrate how one can estimate the
atom-dimer relaxation rate in the case of a wide resonance, for
example in $s$-wave collisions. For an atom and a molecule in a unit
volume, the probability of finding them within the distance $a$ from
each other equals $a^3$ (we assume that there is no $s$-wave
atom-dimer resonance). At distances smaller than $a$, the three-body
wavefunction (in the center-of-mass reference frame) factorizes into
$\Psi({\bf R_1},{\bf R_2},{\bf r})\propto
\rho^{\nu_s-1}\Phi(\hat\Omega)$, where the hyperradius is
$\rho=\sqrt{({\bf R_1-R_2})^2+m_\down/(2m_\down+m_\up)(2{\bf
    r-R_1-R_2})^2}$, $\hat\Omega$ is a five-dimensional set of all the
remaining coordinates (hyperangles), and $\Phi$ is a normalized
hyperangular wavefunction. The power $\nu_s$ for the
$\down\down\up$-system is given by the root of the
transcendental equation \cite{Efimov1973}
\begin{equation}\label{nu_s}
(\nu_s+1) \tan \frac{\pi \nu_s}{2}-2\frac{\sin [\phi (\nu_s+1)]}{\sin (2\phi) \cos (\pi\nu_s/2)}=0
\end{equation}
in the interval $-1 < {\rm Re}\, \nu_s < 3$. In Eq.~(\ref{nu_s})
$\phi$ is defined as $\phi=\arcsin
\left[m_\up/(m_\up+m_\down)\right]$. For $m_\up/m_\down =1$ we obtain
$\nu_s\approx 1.166$ and for $m_\up/m_\down =6.64$ we get
$\nu_s\approx 2.02$.

The probability of finding the three atoms at hyperradii smaller than
$\rho$ scales with $\rho$ as $P(\rho)\propto|\rho^{\nu_s
  -1}|^2\rho^6$. The last term is the volume factor of the
six-dimensional configurational space of the three-body problem (in
the center-of-mass reference frame). Thus, the total probability of
finding the three atoms in the relaxation region is $\sim
a^3 P(R_e)/P(a) \approx a^3 (R_e/a)^{2\nu_s+4}$. The relaxation rate
constant is obtained by multiplying this probability by the frequency
with which the relaxation process takes place once the atoms are
within the range of the potential. It can be estimated as $\sim
1/m_\down R_e^2$. Finally, for the rate constant we obtain
\begin{equation}\label{SWaveRelRateWideRes}
\alpha_s^{\rm ad}\sim \frac{R_e}{m_\down}\left(\frac{R_e}{a}\right)^{2\nu_s+1}. 
\end{equation}

In the case of a narrow resonance the three atoms can approach the
recombination region either as free atoms or as a closed-channel
molecule and an atom. One can show, however, that in this case the probability of the former is much smaller than the
probability of the latter. Indeed, let us consider two atoms, $\up$
and $\down$, in the center-of-mass reference frame. The state of the
system is given by
\begin{equation}\label{state}
|\Psi\rangle=\left( \sum_{\bf k} \psi_{\bf k} \hat a_{{\bf k},\up}^\dagger \hat a_{{\bf -k},\down}^\dagger+ \phi_0 \hat b_0^\dagger\right) |0 \rangle.
\end{equation}
Demanding that Eq.~(\ref{state}) be an eigenstate of the Hamiltonian
(\ref{eq:hamiltonian}) with energy $E$ we get two coupled equations
for $\psi_{\bf k}$ and $\phi_0$, one of which in coordinate space
reads
\begin{equation}\label{OneOfCoupledEquations}
-\frac{\nabla_{\bf R}^2}{2\mu}\psi({\bf R}) + g\phi_0\delta({\bf R})=E\psi({\bf R}).
\end{equation}
From Eq.~(\ref{OneOfCoupledEquations}) one can see that the
singularity of $\psi({\bf R})$ at the origin is related to $\phi_0$ by
\begin{equation}\label{Singularity}
\psi({\bf R}\rightarrow 0)=\phi_0/\sqrt{4\pi R^*}R,
\end{equation}
where we have used the second of Eqs.~(\ref{Parameters}) to express
$g$ in terms of $R^*$. 

Using Eq.~(\ref{Singularity}) it is straightforward to show that the wavefunction of the weakly bound molecular state is given by
\begin{equation}\label{Normalizedphi0}
\phi_{0,\rm b}=\sqrt{Z}=\sqrt{1-1/\sqrt{1+4R^*/a}}
\end{equation}
and
\begin{equation}\label{Normalizedpsi}
\psi_{\rm b}(R)=\sqrt{1-Z}\sqrt{\kappa/2\pi}\exp(\kappa R)/R,
\end{equation}
where $\kappa=\sqrt{2\mu \epsilon_0}$ [see Eq.~(\ref{eq:eb})] and $Z$ is defined in Eq.~(\ref{AppendixZ}). We see that the probability of finding the atoms in the open channel equals $1-Z$ and is small in the regime of intermediate detuning. Therefore, as $R^*/a \rightarrow \infty$ the relaxation rate constant tends to a constant value \cite{remark}
\begin{equation}\label{SWaveBareRateConst}
\alpha_{s,\rm bare}^{\rm ad}\sim R_e/m_\down,
\end{equation}
which corresponds to the relaxation in collisions of atoms and bare molecules.

Equation (\ref{Singularity}) can be used in a more general situation as it applies to a pair of atoms when they are very close to each other. Even in a system of more than two atoms and/or in an external potential we can look at a particular pair of atoms and observe that the probability of finding them in the open channel at separations smaller than $R$ equals $\int_0^R|\psi({\bf R'})|^2 4\pi R'^2 d R'=|\phi_0|^2 R/R^*$, i.e. in the case $R\ll R^*$ it is much smaller than the probability of finding them in the closed channel. In particular, we can conclude that locally, when three atoms are at the hyperradius $\rho \ll R^*$, they can be considered as an atom and a closed-channel molecule. The bare interaction between them is neglected in Eq.~(\ref{eq:hamiltonian}) as it is assumed non-resonant. Their induced interaction (via the exchange of the open-channel atoms) has a Coulomb form \cite{Petrov2004a,WangIncaoEsry2009} and can also be neglected at very small distances.

In order to estimate $\alpha_s^{\rm ad}$ for narrow resonances in the regime of small detuning, $R^*\ll a$, we should slightly modify the speculations that lead us to Eq.~(\ref{SWaveRelRateWideRes}). At distances $\rho \gg R^*$ the three-body wavefunction behaves practically in the same manner as in the wide resonance case. The deviation is important at distances smaller than $R^*$, where, as we have just mentioned, the three-body wavefunction describes a non-interacting atom and a bare molecule. The rate constant reads
\begin{equation}\label{SWaveRelRateNarrowRes}
\alpha_s^{\rm ad} \sim \left[\alpha_{s,\rm bare}^{\rm ad} \left(\frac{1}{R^*}\right)^3\right]\times \left[\left(\frac{R^*}{a}\right)^{2\nu_s+4} a^3\right], 
\end{equation}
where the first factor is the relaxation rate for an atom and a bare molecule confined to a volume of size $R^{*3}$ and the second factor is the probability to find three atoms in this volume. We see that the ratio $\eta_s=\alpha_s^{\rm ad}/\alpha_{s,\rm bare}^{\rm ad}$ interpolates between $\eta_s\sim (R^*/a)^{2\nu_s+1}$ for small $R^*/a$ to $\eta_s=1$ for large $R^*/a$.

The atom-molecule relaxation rate in the $p$-wave channel can be estimated in the same fashion. The difference from the $s$-wave case is the additional factor $(ka)^2$, where $k$ is the relative atom-molecule momentum. It enters due to the unit angular momentum when we calculate the probability to find the atom and the molecule at distances $\sim a$. For the same reason the relaxation rate constant in the collision of an atom and a bare molecule is now momentum dependent, i.e. 
\begin{equation}\label{PWaveBareRateConst}
\alpha_{p,\rm bare}^{\rm ad}(k)\sim R_e^3k^2/m_\down.
\end{equation} 
We also have to take into account the angular momentum when we calculate the relaxation rate of an atom and a bare dimer confined to distances $~R^*$. It now equals $\sim \alpha_{p,\rm bare}^{\rm ad}(k)/k^2 R^{*5}$, where we use the fact that $\alpha_{p,\rm bare}^{\rm ad}(k)/k^2$ is momentum independent in the ultracold limit, $kR_e \ll 1$. One can show that the ratio $\eta_p=\alpha_p^{\rm ad}(k)/\alpha_{p,\rm bare}^{\rm ad}(k)$ should behave as $(R^*/a)^{2\nu_p-1}$ for small $R^*/a$ and should tend to 1 for large $R^*/a$. The power $\nu_p$ is given by the root of the equation \cite{Petrov2003} 
\begin{equation}\label{nu_p}
\frac{\nu_p (\nu_p+2)}{\nu_p+1} \cot \frac{\pi \nu_p}{2}+\frac{\nu_p\sin\phi \cos [\phi (\nu_p+1)]-\sin(\nu_p\phi)}{(\nu_p+1)\sin^2 \phi\cos\phi \sin (\pi\nu_p/2)}=0
\end{equation}
in the interval $-1 < {\rm Re}\, \nu_p < 2$. For $m_\up/m_\down =1$ Eq.~(\ref{nu_p}) gives $\nu_p\approx 0.773$ and for $m_\up/m_\down =6.64$ we get $\nu_p\approx 0.198$. 

\begin{figure}
\includegraphics[width=.48\textwidth]{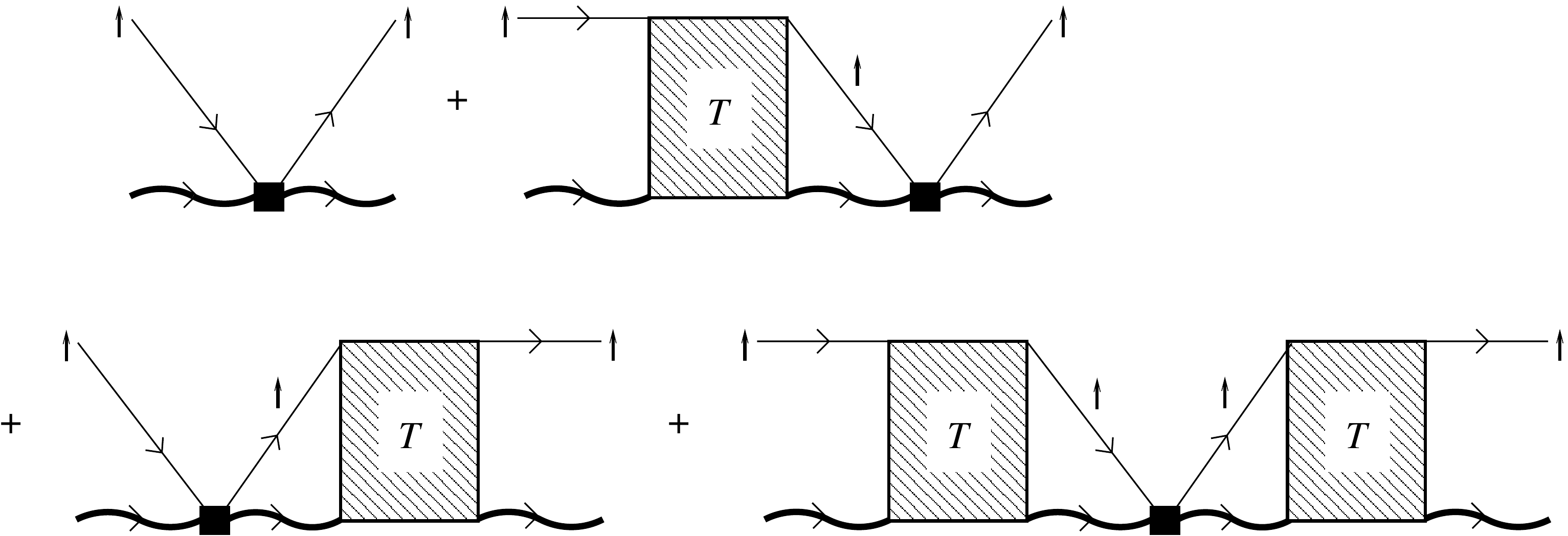}
\caption{The diagrams leading to the first order corrections to the atom-dimer $T$-matrix, Eq. (\ref{eq:dTad}).}
\label{fig:adbare}
\end{figure}

The qualitative analysis of the relaxation rates presented above is valid in the limits of small or large detunings. However, from the practical viewpoint the most interesting is the crossover region, $R^*/a\sim 1$. To calculate the inelastic rates in the general case we add to the Hamiltonian (\ref{eq:hamiltonian}) a weak imaginary short-range interaction potential between the heavy atoms and bare molecules, 
\begin{equation}
H_{\mbox{\tiny ad}}'=-i\sum_{{\bf Q},\k,\p}\frac{\Delta^{\mbox{\tiny ad}}_s + 3\Delta^{\mbox{\tiny ad}}_p\k\cdot \p}{\sqrt{V}}b^\dag_{\p}a^\dag_{\up,{\bf Q}-\p}b_\k a_{\up,{\bf Q}-\k},
\label{eq:Hprime}
\end{equation}
where the parameters $\Delta^{\mbox{\tiny ad}}_s$ and $\Delta^{\mbox{\tiny ad}}_p$ are chosen such that in the extreme limit $R^*/a\gg 1$ the corresponding relaxation rate constants tend to their bare values $\alpha_{s,\rm bare}^{\rm ad}=2 \Delta^{\mbox{\tiny ad}}_s$ and $\alpha_{p,\rm bare}^{\rm ad}(k)= 6 \Delta^{\mbox{\tiny ad}}_p k^2$. 

We can now calculate the ratios $\eta_s$ and $\eta_p$ for finite atom-dimer collision energy and for arbitrary $R^*/a$ by treating Eq.~(\ref{eq:Hprime}) as a perturbation to the Hamiltonian (\ref{eq:hamiltonian}). In Fig.~\ref{fig:adbare} we use the unperturbed (elastic) $T$-matrix found in Sec.~\ref{sec:3body} to construct the first order digrams contributing to the inelastic correction $\delta T$. The explicit on-shell expression reads
\begin{eqnarray}
\delta T_s(k) & = & -iZ\Delta^{\mbox{\tiny ad}}_s\left|\frac{1+\gamma_s(k)}
{1-i\tan\delta_s(k)}\right|^2 \equiv -i\Delta^{\mbox{\tiny ad}}_s\eta_s(k) \nn \\
\delta T_p(k) & = & -iZ\Delta^{\mbox{\tiny ad}}_pk^2\left|\frac{1+\gamma_p(k)}
{1-i\tan\delta_p(k)}\right|^2 \equiv -i\Delta^{\mbox{\tiny ad}}_pk^2\eta_p(k), \nn \\ &&
\label{eq:dTad}
\end{eqnarray}
where we define
\begin{eqnarray}
\gamma_s(k) & = & \frac2\pi{\cal P}\int q^2dq\frac{K_s(k,q)}{q^2-k^2} \\
\gamma_p(k) & = & \frac2{\pi k}{\cal P}\int q^3dq\frac{K_p(k,q)}{q^2-k^2}.
\end{eqnarray}
These integrals should be performed by taking the principal value, and $K_\ell(k,q)=\tilde f_\ell(k,q)/[1+ik\tilde f_\ell(k,k)]$ is the solution of Eq.~(\ref{eq:inteq}) obtained by using the principal value prescription.

\begin{figure}
\includegraphics[width=.5\textwidth]{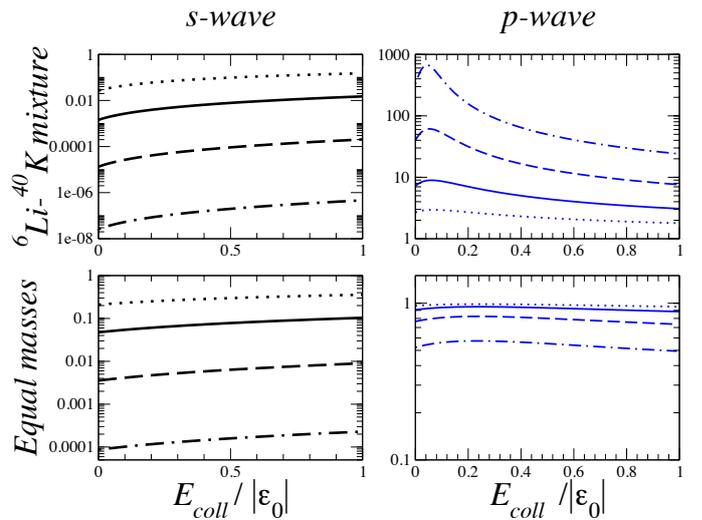}
\caption{(color online). The ratios $\eta_s$ (left) and $\eta_p$ (right) versus collision energy for various detunings: $R^*/a=4$ (dotted), 1 (solid), 1/4 (dashed), and 1/16 (dot-dashed).}
\label{fig:ratesvsecol}
\end{figure}

\begin{figure}
\includegraphics[width=.5\textwidth]{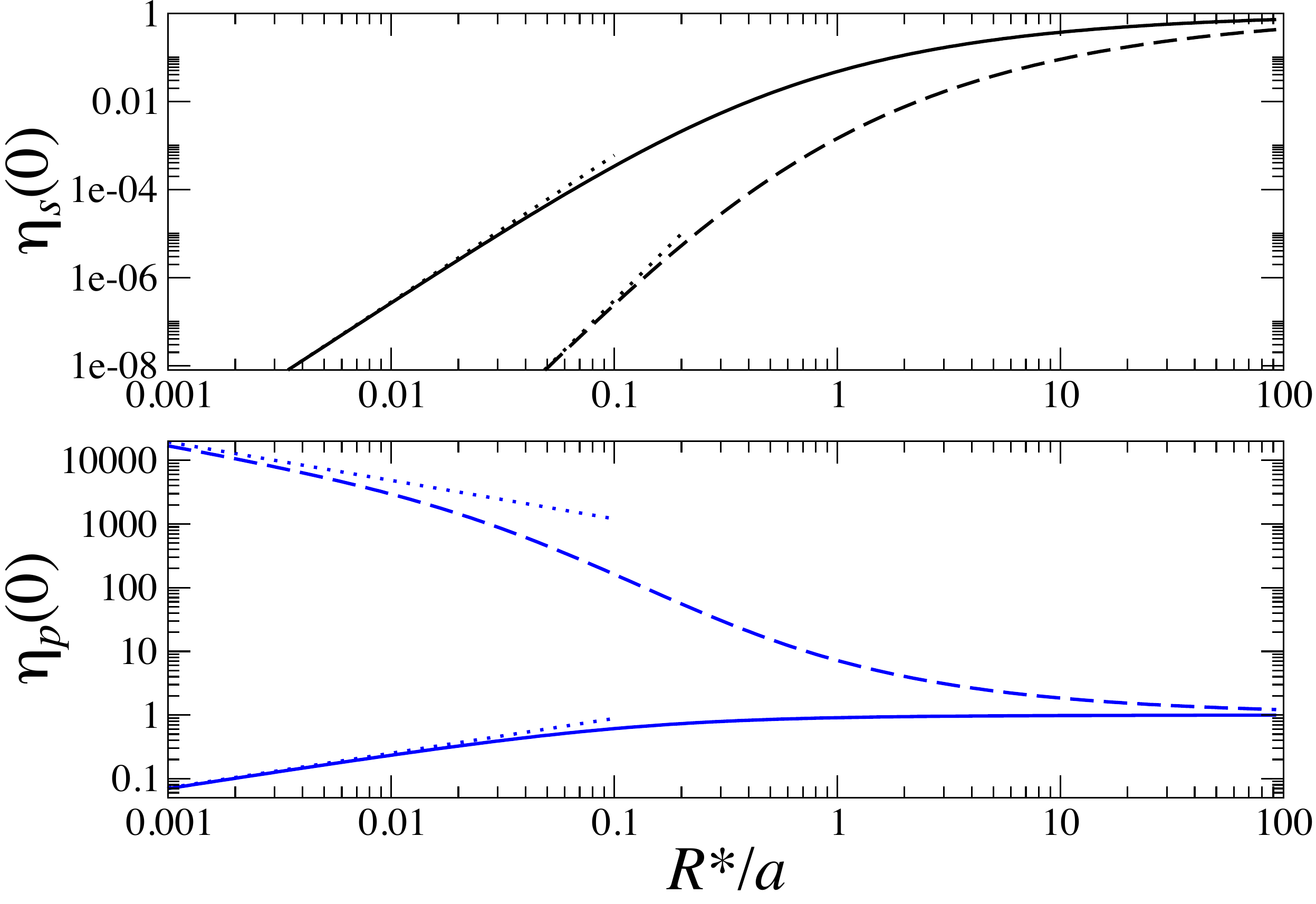}
\caption{(color online). The ratios $\eta_s$ (top) and $\eta_p$ (bottom) at zero
  collision energy versus $R^*/a$ for the equal mass case (solid) and
  the Li-K mixture (dashed). We also show with dotted lines the
  corresponding power scalings in the limit of small $R^*/a$ (see
  text). }
\label{fig:alphasp}
\end{figure}

Results of our calculation are summarized in Figs.~\ref{fig:ratesvsecol} and \ref{fig:alphasp}. In Fig.~\ref{fig:ratesvsecol} we present the collision energy dependence of $\eta_s$ and $\eta_p$, and Fig.~\ref{fig:alphasp} shows their zero energy values versus the detuning $R^*/a$. According to their definition these parameters are the suppression or enhancement factors of the relaxation rate constants $\alpha_{s}^{\rm ad}$ and $\alpha_{p}^{\rm ad}(k)$ compared to their bare values $\alpha_{s,\rm bare}^{\rm ad}$ and $\alpha_{p,\rm bare}^{\rm ad}(k)$. Note that in the ultracold limit the $s$-wave relaxation is much more dangerous than the $p$-wave one [compare Eqs.~(\ref{SWaveBareRateConst}) and (\ref{PWaveBareRateConst})]. Figure~\ref{fig:alphasp} thus suggests that the atom-dimer collisions in the heteronuclear case are much less prone to relaxation than in the homonuclear one. Indeed, in the Li-K case already for $R^*/a=1$ the $s$-wave relaxation is suppressed by three orders of magnitude. We think that this is promising for the system's longevity even though the $p$-wave relaxation for this $R^*/a$ is enhanced by an order of magnitude. At this point we can also make the following observation: in order to suppress collisional relaxation in the homonuclear case one should make the detuning as small as possible, whereas in the Li-K case there exists an optimal detuning where the inelastic atom-dimer collisional losses reach their minimum. The exact value of such optimal detuning depends on the average momentum $k$ and on the actual values of the bare relaxation rates $\alpha_{s,\rm bare}^{\rm ad}$ and $\alpha_{p,\rm bare}^{\rm ad}(k)$.

\subsection{Relaxation in dimer-dimer collisions}

\begin{figure}
\includegraphics[width=.48\textwidth]{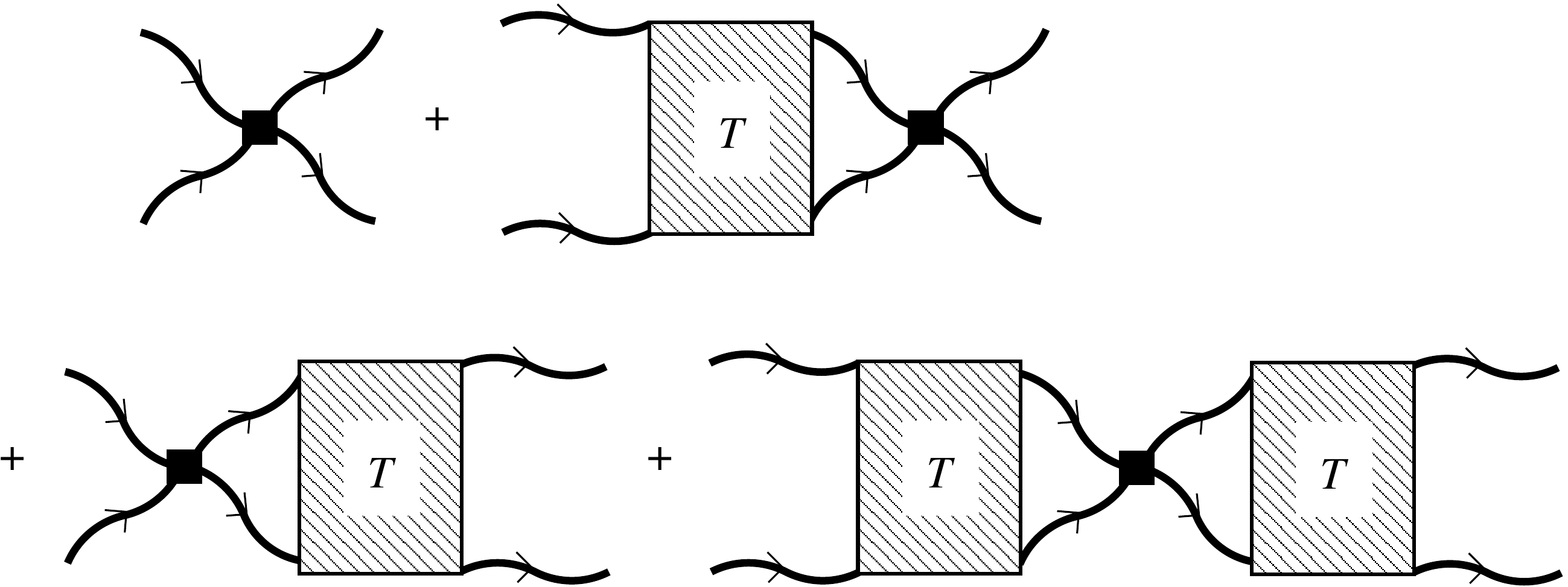}
\caption{The diagrams leading to the first order correction to the dimer-dimer $T$-matrix, Eq. (\ref{eq:dTdd}).}
\label{fig:ddbare}
\end{figure}

\begin{figure}
\includegraphics[width=.48\textwidth]{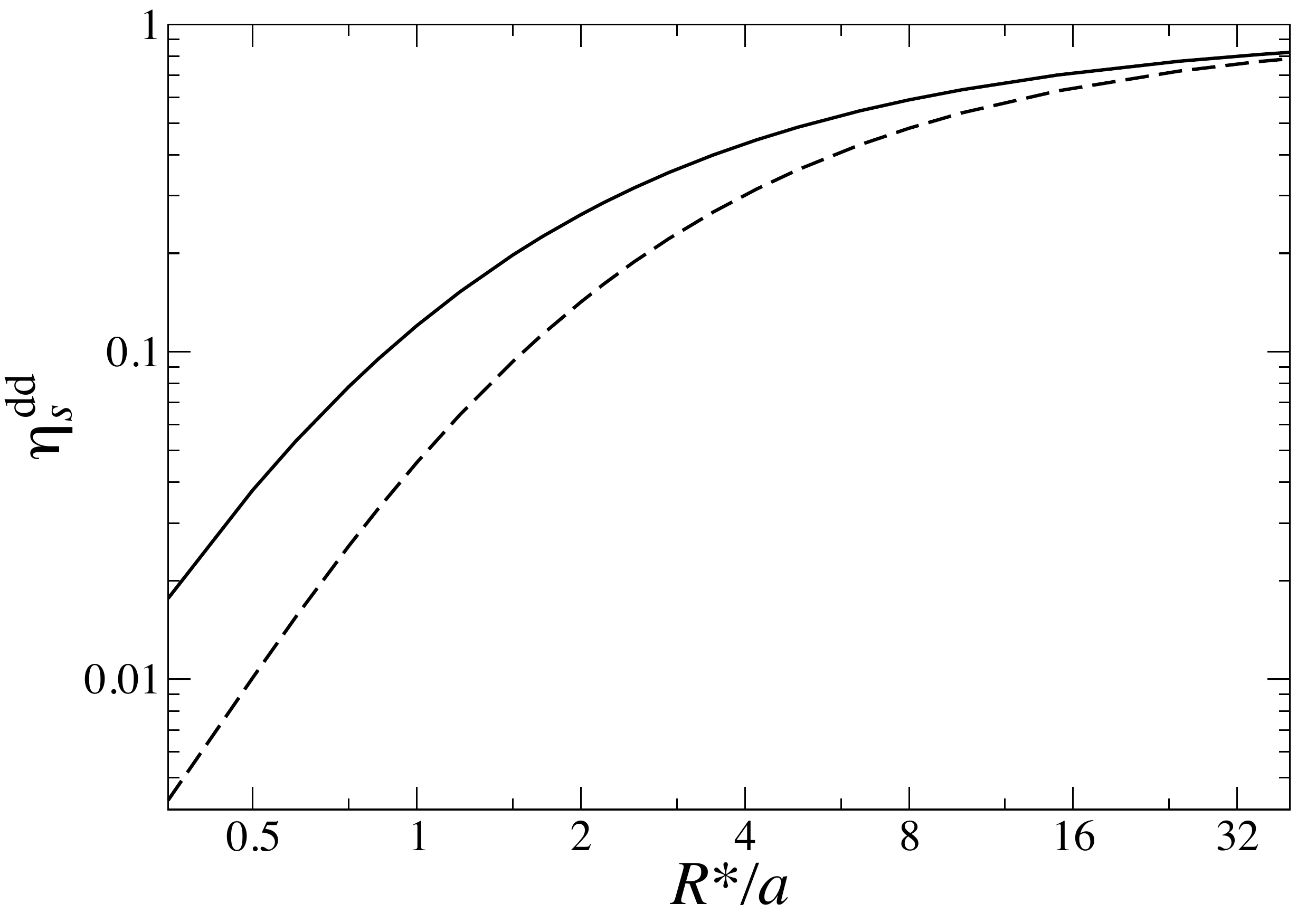}
\caption{The relaxation rate in dimer-dimer collisions.}
\label{fig:alpha4}
\end{figure}

The dimer-dimer scattering problem is more complicated than the atom-dimer one in many respects. In particular, one has to deal not only with the relaxation channel which requires two heavy and one light fermion to approach each other to short distances but also with analogous processes in the heavy-light-light subsystem. In fact, in the regime of intermediate detuning ($R^*/a\gg 1$) the occupation of the bare molecular states is dominant, and the above mentioned ``three-body'' channels are suppressed as they require one of the atoms to be in the open channel. Then the dominant decay scenario is the relaxation in collisions of bare molecules. In this case the perturbation is
\begin{equation}
H_{\mbox{\tiny dd}}'=-i\sum_{{\bf Q},\k,\p}\frac{\Delta^{\mbox{\tiny dd}}_s}{\sqrt{V}}b^\dag_{\p}b^\dag_{{\bf Q}-\p}b_\k b_{{\bf Q}-\k},
\label{eq:Hdd}
\end{equation}
and we treat it in the same fashion as (\ref{eq:Hprime}). Equation~(\ref{eq:Hdd}) correctly describes the relaxation in the limit $R^*/a\gg 1$ if we set $\Delta^{\mbox{\tiny dd}}_s=\alpha^{\rm dd}_{\rm bare}/4$. The diagrams leading to the lowest order correction to the dimer-dimer $T$-matrix are depicted in Fig.~\ref{fig:ddbare}. The on-shell $\delta T$ at vanishing collision energy equals
\begin{eqnarray}
\delta T(0)  & = & -2iZ^2\Delta^{\mbox{\tiny dd}}_s\left|1+\frac{i}{g^4Z^2}\int 
\frac{d^4q}{(2\pi)^4}D(\q,\epsilon_0+q_0)\right.\nn \\ &&
\hspace{25mm} \times \left.D(-\q,\epsilon_0-q_0)
T(q,q_0)\vphantom{\int}\right|^2 \nn \\ 
& \equiv & -2i\Delta^{\mbox{\tiny dd}}_s\eta_s^{\mbox{\tiny dd}}, 
\label{eq:dTdd}
\end{eqnarray}
where $T(q,q_0)$ is the solution of the dimer-dimer integral equation
(\ref{eq:TGamma}), and $\eta_s^{\mbox{\tiny dd}}$ is the ratio of the
actual relaxation rate constant at a given $R^*/a$ to its bare value
(at $R^*/a=\infty$). This parameter is plotted in
Fig.~\ref{fig:alpha4} versus the detuning $R^*/a$ and we see that this
``four-atom'' relaxation mechanism is suppressed for small
detunings. The power law dependence in this limit can be understood in
a similar fashion as in the atom-dimer case. We start with two
molecules in a unit volume. The probability to find them within the
distance $\sim a$ equals $a^3$. Then, at shorter distances the
four-body wavefunction is proportional to $\rho^{\nu_{\rm 4body}-1}$,
where $\rho$ is the four-body hyperradius. Thus, given that the four
atoms are confined to the volume $\sim a^3$, the probability to find
them at hyperradii $\rho\lesssim R^*$ equals $(R^*/a)^{2\nu_{\rm
    4body}+7}$, where we take into account that the four-body
configurational volume in the center-of-mass frame scales as
$\rho^9$. At hyperradii $\rho\lesssim R^*$ we deal with two bare
molecules and the relaxation rate is given by $\alpha^{\rm dd}_{\rm
  bare}/R^{*3}$. Finally, for the dimer-dimer relaxation rate constant
associated with the ``four atom'' relaxation mechanism in the limit
$R^*/a\ll 1$ we obtain
\begin{equation}\label{DimerDimerRelRateNarrowRes}
\alpha^{\rm dd}=\alpha^{\rm dd}_{\rm bare}(R^*/a)^{2\nu_{\rm 4body}+4}. 
\end{equation}
The power $\nu_{\rm 4body}$ may be obtained by calculating the ground
state energy of the four-body system in the unitarity limit in a
harmonic potential which has been carried out in Ref.~\cite{Blume} for
various mass ratios. We cite the following results \cite{Blume}:
$\nu_{\rm 4body}\approx 0.0$ for $m_\up/m_\down=1$, $\nu_{\rm
  4body}\approx 0.3$ for $m_\up/m_\down=4$, and $\nu_{\rm
  4body}\approx 0.5$ for $m_\up/m_\down=8$. Our numerical calculations
for the dimer-dimer relaxation in the limit $R^*/a\ll 1$ are
consistent with Eq.~(\ref{DimerDimerRelRateNarrowRes}) with these
powers.

Figure~\ref{fig:alpha4} only shows the result corresponding to the ``four-body'' mechanism of dimer-dimer relaxation (dominant at intermediate detunings) and thus presents a lower bound for the relaxation rate constant. We see that for $R^*/a\sim 1$ the dimer-dimer relaxation is less suppressed compared to the atom-dimer case. Consequently, this region of detunings is more suitable for studies of atom-dimer mixtures with low concentration of molecules. Otherwise, if one wants to study mixtures with higher molecular concentrations, it is necessary to decrease the detuning as much as possible.

In the regime of small detuning the ``three-body'' mechanisms of
dimer-dimer relaxation can be as important as the ``four-body''
one. For wide resonances the ``three-body'' cases are described in
detail in Ref.~\cite{Petrov2005a,Petrov2005}. These results can be
easily generalized to the regime of small detunings near a narrow
resonance by counting probabilities as we did earlier in this section
in order to estimate the atom-dimer relaxation rates. Let us present the
final results. The dimer-dimer relaxation rate constant originating
from the $s$-wave atom-dimer relaxation mechanism equals (up to a
numerical prefactor) the one for the atom-dimer collisions given by
Eq.~(\ref{SWaveRelRateNarrowRes}). Namely,
\begin{equation}\label{SWaveDimerDimerRelRateNarrowRes}
\alpha^{\rm dd}_{s - {\rm ad}} \sim \alpha_{s,\rm bare}^{\rm ad} (R^*/a)^{2\nu_s+1}.
\end{equation}
The dimer-dimer relaxation rate constant originating from the $p$-wave
atom-dimer relaxation mechanism equals
\begin{eqnarray}
\alpha^{\rm dd}_{p - {\rm ad}} & \sim & \alpha_{p,\rm bare}^{\rm
  ad}(1/a) (R^*/a)^{2\nu_p-1}
\nn \\ & \sim & (R_e^3/m_\down a^2)(R^*/a)^{2\nu_p-1}.
\label{PWaveDimerDimerRelRateNarrowRes}
\end{eqnarray}
Equations (\ref{SWaveDimerDimerRelRateNarrowRes}) and (\ref{PWaveDimerDimerRelRateNarrowRes}) can be modified to describe the light-light-heavy relaxation mechanism. In this case the bare rates $\alpha_{s,\rm bare}^{\rm ad}$ and $\alpha_{p,\rm bare}^{\rm ad}(k)$ should be taken for the collisions of light atoms and bare molecules, and the powers $\nu_s$ and $\nu_p$ should correspond to the light-light-heavy three-body system with zero and unit angular momenta respectively. For $m_\up/m_\down=m_{\rm K}/m_{\rm Li}$ they equal $\nu_s\approx 1.01$ and $\nu_p\approx 0.945$.

\section{Concluding remarks \label{sec:conc}}

Concluding the paper, we would like to discuss several issues, which,
from our viewpoint, are important for future studies. We believe that
the $p$-wave atom-dimer resonance should significantly influence the
behavior of a mixture of weakly bound K-Li molecules and K atoms. The
``simplest'' many-body problem in which the effect of the resonance
should be visible is the problem of a single atom immersed in a BEC of
molecules or a single molecule immersed in a Fermi sea of atoms. Let
us consider, for example, the former case. In the limit $v_{\rm ad}n\ll
1$, where $n$ is the BEC density, we compute the correction to the dispersion
curve of the polaron due to the $p$-wave interaction with
molecules. By using the standard diagrammatic techniques, the
dispersion is found to be
\begin{equation}\label{MFDispersion}
  \epsilon_K(q)\simeq \frac{q^2}{2m_\up} + \frac{3\pi n v_{\rm ad} q^2}{2\mu_3}.
\end{equation}
We see that in the case of a negative and large $p$-wave scattering
volume $v_{\rm ad}$, the effective mass of the polaron increases, which
can be measured in an experiment similar to the one performed recently
at ENS on collective modes of a homonuclear $^6$Li spin-mixture
\cite{Nascimbene2009}. Most interesting seems to be the regime
$|v_{\rm ad}|n\sim 1$, where the validity of the mean-field approach is not
guaranteed even on the qualitative level and a non-perturbative
analysis is required. Our few-body results suggest that the crossover
phase diagram in the K-Li case should be less ``mean-field friendly''
than the one in the homonuclear case. The same holds for the case of a
single heavy-light molecule immersed in a Fermi sea of the heavy
species. Variational analysis \cite{Mathy2010} suggests that such
molecules can form a BEC with finite momentum (an analog of the FFLO
phase). In a related work \cite{Radzihovsky2009}, it was demonstrated
that the phase diagram of a two-species bosonic mixture near a
$p$-wave interspecies Feshbach resonance is quite rich and features,
in particular, a finite-momentum superfluid phase.

In this paper we have shown that the atom-dimer scattering is
characterized by a very strong dependence on the angular momentum, the
effect being remarkably pronounced for the K-Li mass ratio. To observe
the interference between the $s$- and $p$-waves we have proposed a
scattering experiment in which one collides a cloud of atoms with a
cloud of molecules. Depending on the collision energy the scattering
goes predominantly in the forward or backward directions. The contrast
in this experiment is limited to approximately $75\%$ due to the fact
that only two partial waves are involved. We believe that a similar
experiment in a quasi-1D geometry would show a complete destructive or
constructive interference, i.e. transmissionless or reflectionless 1D
scattering at certain values of the collision energy. The control over
the atom-dimer scattering properties provided by the external quasi-1D
confinement opens up interesting perspectives for observing rather
exotic effects. For example, in the case of a vanishing transmission
coefficient the propagation of a molecule immersed in a quasi-1D gas
of heavy fermions is suppressed, leading to degeneracies in its
excitation spectrum and an unusual dynamical correlation function
\cite{Zvonarev2007}. Besides, as in the quasi-2D case
\cite{Levinsen2009}, the quasi-1D confinement can push the trimer state
below the atom-dimer threshold, i.e. make it bound. The quasi-1D
confinement of only one of the species leads to similar effects and
can be used to control the atom-dimer scattering amplitude
\cite{Tan2008,Tan2009}.

In the present paper we focus on the $^6$Li-$^{40}$K mixture in which
the Feshbach molecule is bosonic and the enhanced $p$-wave interaction
occurs between a boson and a fermion. On the other hand, our approach
applies equally well to the case of light bosonic atoms. If, for
example, the light particle is chosen to be $^7$Li, the enhanced
interaction takes place between a fermionic atom and a fermionic
dimer. In this case, however, relaxation in molecule-molecule
collisions is not suppressed by the Pauli exclusion principle as two
bosons and one fermion can easily approach each other. Nevertheless,
the fact that the molecules are identical fermions is advantageous for
the system's longevity. Such a mixture of fermionic atoms and
molecules can be seriously considered for studies of the BCS-BEC
crossover with resonant $p$-wave interspecies interactions.

\begin{acknowledgement}

  We acknowledge support by the EuroQUAM-FerMix program, by the IFRAF
  Institute, and by the Russian Foundation for Fundamental
  Research. One of us (J.L.) acknowledges support by a Marie Curie
  Intra European Fellowship within the 7th European Community
  Framework Programme. LPTMS is a mixed research unit No. 8626 of CNRS
  and Universit\'e Paris Sud.

\end{acknowledgement}

\appendix

\section{Two-body scattering amplitude and propagator of
  dimers \label{app:a}}

\begin{figure}[tb]
\includegraphics[width=\hsize,clip]{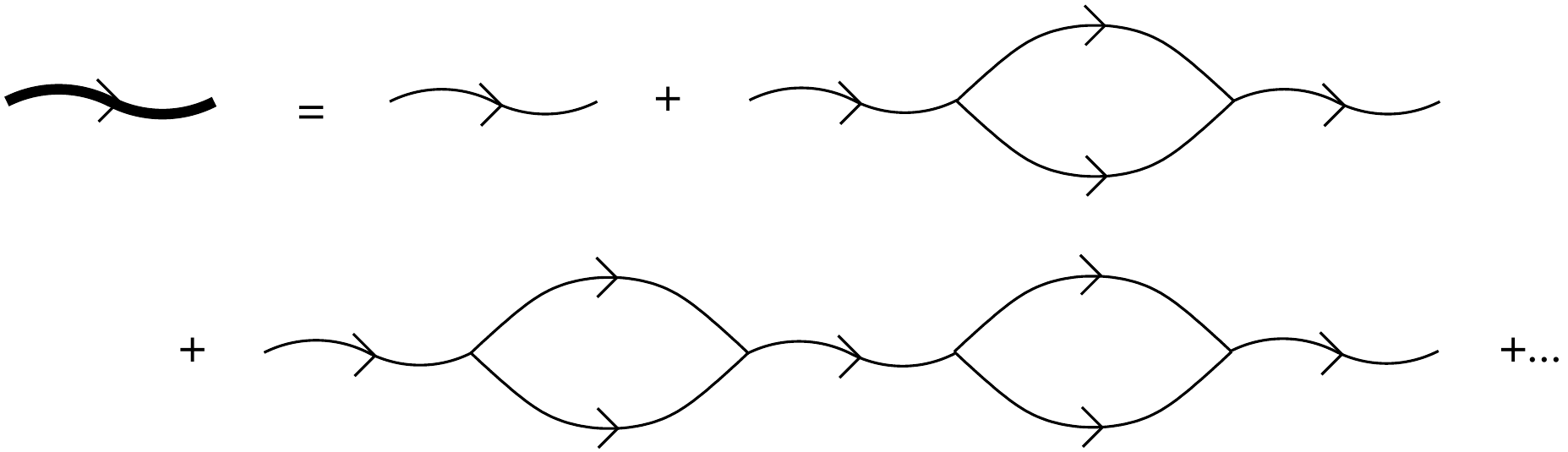}
\caption{The propagator of dimers consists of an admixture of closed
  channel Feshbach molecules (thin wave lines) and open channel
  fermionic loops.}
\label{fig:bosonprop}
\end{figure}
In this appendix we obtain the dressed dimer propagator and relate the
bare parameters of the Hamiltonian, Eq. (\ref{eq:hamiltonian}), to the
physical observables of the scattering process between a heavy and a
light fermionic atom. The propagator of the bare (closed-channel)
molecule is
\begin{equation}
D_0(\p,p_0) = \frac1{p_0-p^2/2M-\omega_0+i0}.
\end{equation}
The dressed propagator is obtained by resumming the diagrams of
Fig. \ref{fig:bosonprop}, resulting in
\begin{equation}
D(\p,p_0) = \frac{g^2}{D_0^{-1}(\p,p_0)-g^2\Pi(\p,p_0)}.
\end{equation}
The fermion loop $\Pi$ is
\begin{eqnarray}
  \Pi(\p,p_0) & \equiv & i\int \frac{d^4Q}{(2\pi)^{-4}}
G_\up(p+Q)G_\down(-Q) \label{eq:dressing} \\
  & = &-\frac{\mu\Lambda}{\pi^2}+\frac{\mu^{3/2}}{\sqrt{2}\pi}\sqrt{-p_0+p^2/2M-i0}.
\end{eqnarray}
The cut-off at large momenta, $\Lambda$, is the inverse van der Waals
range of the potential. We have chosen to include an extra factor
$g^2$ in the dressed dimer propagator. This is purely for bookkeeping
purposes, as this propagator always appears along with a factor of
$g^2$ in Feynman diagrams, providing an easy manner of keeping tracks
of powers of the coupling constant. The dressed propagator is
\begin{eqnarray}
D(\p,p_0) & = & g^2\left[p_0-\frac{p^2}{2M}-\omega_0+\frac{g^2\mu\Lambda}{\pi^2}
\vphantom{\sqrt{-p_0+\frac{p^2}{2M}-i0}}\right. \nn \\
&&
\left.-\frac{g^2\mu^{3/2}}{\sqrt{2}\pi}\sqrt{-p_0+\frac{p^2}{2M}-i0}\right]^{-1}\hspace{-1mm}.
\label{eq:ddressedapp}
\end{eqnarray}

\begin{figure}[tb]
\includegraphics[width=\hsize,clip]{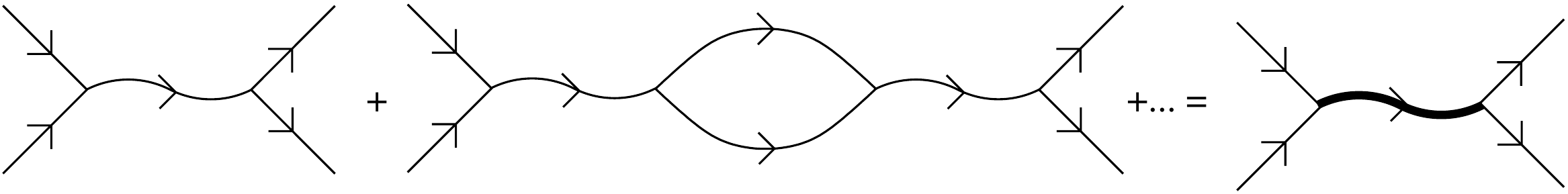}
\caption{The relationship between the scattering amplitude for elastic
  heavy-light atom scattering and the dressed dimer propagator.}
\label{fig:t2}
\end{figure}

Let us now relate the bare parameters, $g$ and $\omega_0$, to the
physical parameters $a$ and $R^*$ of Eq. (\ref{eq:delta0lowk}). To this
end we calculate the amplitude of elastic heavy-light atom scattering
in the model \rfs{eq:hamiltonian}. Let the heavy atom have
four-momentum $(\k,k^2/2m_\up)$ and the light have
$(-\k,k^2/2m_\down)$. The scattering is depicted in
Fig. \ref{fig:t2}. The summation of diagrams takes the same form as
above, Eq. (\ref{eq:dressing}), and recalling that $g^2$ has been
absorbed into the dimer propagator, the scattering amplitude is
\begin{eqnarray}
T_2(\k,-\k) & = & D(\0,k^2/2\mu) \nn \\
& = & \frac{2\pi}\mu\frac1{\frac{2\pi}{\mu g^2}
\left(-\omega_0+\frac{g^2\mu\Lambda}{\pi^2}\right)+\frac{\pi}{\mu^2 g^2}k^2+ik}.
\nn \\
\end{eqnarray}
Comparing with the scattering amplitude at low momenta, $k\lesssim
a^{-1}$,
\begin{equation}
T_2(\k,-\k) \equiv -\frac{2\pi}\mu f(\k,-\k)\approx
\frac{2\pi}\mu\frac1{a^{-1}+R^* k^2+ik}
\end{equation}
the scattering length and effective range are found to be
\begin{equation}
a=\frac{\mu g^2}{2\pi}\frac1{\frac{g^2\mu\Lambda}{\pi^2}-\omega_0}, 
\hspace{1cm} R^*=\frac{\pi}{\mu^2g^2}.
\end{equation}
In terms of the physical observables, the dimer propagator
(\ref{eq:ddressedapp}) then takes the form
(\ref{eq:DressedMoleculePropagator}).
%\begin{widetext}
%\begin{equation}
%D(\p,p_0) = \frac{2\pi}\mu\frac1{2\mu R^*\left(p_0-\frac{p^2}{2M}+i0\right)
%+a^{-1}-\sqrt{2\mu}\sqrt{-p_0+\frac{p^2}{2M}-i0}}.
%\end{equation}
%\begin{eqnarray}
%D(\p,p_0) & = & \nn \\ && \hspace{-19mm}
%\frac{2\pi}\mu\frac1{2\mu R^*\left(p_0-\frac{p^2}{2M}+i0\right)
%+a^{-1}-\sqrt{2\mu}\sqrt{-p_0+\frac{p^2}{2M}-i0}}. \nn \\
%\end{eqnarray}
%\end{widetext}

The binding energy of the heteronuclear molecule is found as the
energy pole of the propagator and is
\begin{equation}
\epsilon_0 = -\frac{\left[\sqrt{1+4R^*/a}-1\right]^2}{8\mu R^{*2}}.
\end{equation}
Since $R^*\geq0$ there is only one such pole on the physical
sheet. The residue at the energy pole is needed for proper
renormalization of the three- and four-body T-matrices, and is
\begin{equation}\label{AppendixZ}
g^2Z = \frac{\pi}{\mu^2R^*}\left(1-\frac1{\sqrt{1+4R^*/a}}\right).
\end{equation}
Each external dimer propagator acquires a factor $\sqrt{Z}$ in the
few-body scattering problems.

\section{The kernel of the 4-body integral equation \label{app:b}}
$\Gamma^{(0)}$ is the result of calculating
the Born diagram,
\begin{eqnarray}
  \Gamma^{(0)}(q,q_0;p,p_0) & = & \nn \\ && \hspace{-30mm}
-i\int\frac{d\Omega_{\vec q}}{4\pi}
  \int\frac{d^4Q}{(2\pi)^4}
  G_\up\left(\Q+\frac{\p}{2}+\frac{\q}2,\frac{\epsilon_0}2+Q_0+
    \frac{p_0}2+\frac{q_0}2\right)
\nn \\ && \hspace{-5mm} \times
  G_\up\left(\Q-\frac{\p}2-\frac{\q}2,\frac{\epsilon_0}2+Q_0-\frac{p_0}2
    -\frac{q_0}2\right)
 \nn \\ && \hspace{-5mm} \times
  G_\down\left(-\Q-\frac{\p}2+\frac{\q}2,\frac{\epsilon_0}2-Q_0-\frac{p_0}2
    +\frac{q_0}2\right)
\nn \\ && \hspace{-5mm} \times
  G_\down\left(-\Q+\frac{\p}2-\frac{\q}2,\frac{\epsilon_0}2-Q_0+\frac{p_0}2-
    \frac{q_0}2\right)
  \nn \\
  && \hspace{-15mm}
= -2\int\frac{d\Omega_{\vec q}}{4\pi}
  \int\frac{d^3Q}{(2\pi)^3}\frac{A}{(A^2-B^2)(A^2-C^2)},
\end{eqnarray}
with
\begin{eqnarray}
  A & = &
  \epsilon_0-\frac{Q^2}{2\mu}-\frac{p^2}{8\mu}-\frac{q^2}{8\mu}
-\frac{\p\cdot \q}{4m_\up}+
 \frac{\p\cdot\q}{4m_\down}, \nn \\
  B & = & p_0-\frac{\Q\cdot\p}{2m_\up}+\frac{\Q\cdot\p}{2m_\down}
  -\frac{\Q\cdot\q}{2\mu}, \nn \\
  C & = & q_0-\frac{\Q\cdot\q}{2m_\up}+\frac{\Q\cdot\q}{2m_\down}
  -\frac{\Q\cdot\p}{2\mu}.
\end{eqnarray}

%\bibliographystyle{epj}
%\bibliography{biblio}

\begin{thebibliography}{}
%Author, \textit{Book title} (Publisher, place year) page numbers
% etc

\bibitem{Inguscio2008}
M. Inguscio, W. Ketterle, C. Salomon, eds.,
\textit{Ultracold Fermi Gases, Proceedings of the International School
  of Physics ``Enrico Fermi''}, (IOS Press, Amsterdam 2008)

\bibitem{Giorgini2008}
S. Giorgini, L.P. Pitaevskii, S. Stringari,
Rev. Mod. Phys. \textbf{80}, 1215  (2008)

\bibitem{Taglieber2008}
M. Taglieber,  A.-C. Voigt, T. Aoki, T.W. H\"ansch, K. Dieckmann, 
Phys. Rev. Lett., \textbf{100}, 010401 (2008)

\bibitem{Wille2008}
E. Wille, F.M. Spiegelhalder, G.Kerner, D. Naik, A. Trenkwalder,
G. Hendl, F. Schreck, R. Grimm, T.G. Tiecke, J.T.M. Walraven,
S.J.J.M.F. Kokkelmans, E. Tiesinga, P.S. Julienne,
Phys. Rev. Lett. \textbf{100}, 053201 (2008)

\bibitem{Voigt2009}
A.-C. Voigt, M. Taglieber, L. Costa, T. Aoki, W. Wieser,
T.W. H\"ansch, K. Dieckmann,
Phys. Rev. Lett. \textbf{102}, 020405 (2009)

\bibitem{Spiegelhalder2009}
F.M. Spiegelhalder, A. Trenkwalder, D. Naik, G. Hendl, F. Schreck, R. Grimm,
Phys. Rev. Lett. \textbf{103}, 223203 (2009)

\bibitem{Tiecke2009}
T.G. Tiecke, M.R. Goosen, A. Ludewig, S.D. Gensemer, S. Kraft,
S.J.J.M.F. Kokkelmans, J.T.M. Walraven, Phys. Rev. Lett. \textbf{104},
053202 (2010)

\bibitem{Spiegelhalder2010}
F.M. Spiegelhalder, A. Trenkwalder, D. Naik, G. Kerner, E. Wille,
G. Hendl, F. Schreck, R. Grimm,
Phys. Rev. A \textbf{81}, 043637 (2010)

\bibitem{Dieckmann2010}
L.~Costa, H.~Brachmann, A.-C.~Voigt, C.~Hahn, M.~Taglieber, T.~W.~Hänsch, K.~Dieckmann, Phys. Rev. Lett. \textbf{105}, 123201 (2010)

\bibitem{Trenkwalder2010}
A. Trenkwalder, C. Kohstall, M. Zaccanti, D. Naik, A.I. Sidorov, F. Schreck, R. Grimm,
arXiv:1011.5192v1 (2010)

\bibitem{note:optlat}
A species selective optical lattice can be applied to modify the effective mass ratio.

\bibitem{Petrov2007}
D.S. Petrov, G.E. Astrakharchik, D.J. Papoular, C. Salomon,
G.V. Shlyapnikov,
Phys. Rev. Lett. \textbf{99}, 130407 (2007)

\bibitem{LL}
L.D. Landau, E.M. Lifshitz, \textit{Quantum Mechanics},
(Butterworth-Heinemann, Oxford 1981)

\bibitem{Efimov1973}
V.N. Efimov, Nucl. Phys. A \textbf{210}, 157 (1973)

\bibitem{Fonseca1979}
A.C. Fonseca, E.F. Redish, P.E. Shanley, Nucl. Phys. A. \textbf{320},
273 (1979)

\bibitem{Amado1971}
R.D. Amado, J.V. Noble, Phys. Lett. \textbf{35B}, 25 (1971)

\bibitem{Petrov2005a}
D.S. Petrov, C. Salomon, G.V. Shlyapnikov,
J. Phys B: At. Mol. Opt. Phys. \textbf{38}, S645 (2005)

\bibitem{Petrov2003}
D.S. Petrov, Phys. Rev. A \textbf{67}, 010703(R) (2003)

\bibitem{Kartavtsev2007}
O.I. Kartavtsev, A.V. Malykh, J. Phys. B \textbf{40}, 1429 (2007)

\bibitem{Levinsen2009}
J. Levinsen, T.G. Tiecke, J.T.M. Walraven, D.S. Petrov,
Phys. Rev. Lett. \textbf{103}, 153202 (2009)

\bibitem{note:manybody}
We note that in a many-body problem, the width of the resonance can be
characterized by comparing $|r_0|$ with the mean interparticle
separation. In particular, the narrow resonance condition
$\left(n|r_0|^3\right)^{-1/3}\ll 1$ (much more strict than $|r_0|\gg R_e$) allows
for a perturbative expansion across the whole BCS-BEC crossover
\cite{Gurarie2007}.

\bibitem{Petrov2004a}
D.S. Petrov, Phys. Rev. Lett. {\bf 93}, 143201 (2004)

\bibitem{Gurarie2007}
V. Gurarie, L. Radzihovsky, Ann. Phys. {\bf 322}, 2 (2007)

\bibitem{note:model}
In principle, the method that we develop in this paper can be
generalized to an arbitrary energy dependence of the phase shift.

\bibitem{Timmermans1999}
E. Timmermans, P. Tommasini, M. Hussein, A. Kerman
Physics Reports {\bf 315}, 199 (1999)

\bibitem{Skorniakov1956}
G.V. Skorniakov, K.A. Ter-Martirosian,
Zh. Eksp. Teor. Phys. {\bf 31}, 775 (1956), [Sov. Phys. JETP {\bf 4}, 648 (1957)]

\bibitem{Bedaque1998}
P.F. Bedaque, H.-W. Hammer, U. van Kolck,
Phys. Rev. C {\bf 58}, R641 (1998)

\bibitem{Born1927}
M. Born, J.R. Oppenheimer, Ann. der Phys. {\bf 84}, 457 (1927)

\bibitem{Bethe1935}
H. Bethe, R. Peierls, Proc. R. Soc. London {\bf A148}, 146, (1935)

\bibitem{note:high}
Quantum interference of $s$- and $d$-waves has been observed in
collisions of $^{87}$Rb BECs at rather high collision energies, see
Ch. Buggle, J. L\'eonard, W. von Klitzing, J.T.M. Walraven,
Phys. Rev. Lett. {\bf 93}, 173202 (2004); 
N.R. Thomas, N.~Kj{\ae}rgaard, P.S. Julienne, A.C. Wilson,
Phys. Rev. Lett. {\bf 93}, 173201 (2004)

\bibitem{Levinsen2006}
J. Levinsen, V. Gurarie, Phys. Rev. A {\bf 73}, 053607 (2006)

\bibitem{Petrov2004}
D.S. Petrov, C. Salomon, G.V. Shlyapnikov
Phys. Rev. Lett. {\bf 93}, 090404 (2004)

\bibitem{Brodsky2005}
I.V. Brodsky, A.V. Klaptsov, M.Yu. Kagan, R. Combescot, X. Leyronas
JETP Lett. {\bf 82}, 273 (2005)

\bibitem{Astrakharchik2004}
G.E. Astrakharchik, J. Boronat, J. Casulleras, S. Giorgini,
Phys. Rev. Lett. {\bf 93}, 200404 (2004)

\bibitem{Stecher2007}
J. von Stecher, C.H. Greene, D. Blume,
Phys. Rev. A {\bf 76}, 053613 (2007)

\bibitem{Petrov2005}
D.S. Petrov, C. Salomon, G.V. Shlyapnikov, Phys. Rev. A
{\bf 71}, 012708 (2005)

\bibitem{Levinsen2007a}
J. Levinsen, Ph.D. thesis, University of Colorado at Boulder (2007),
arXiv:0807.2840v1

\bibitem{Marcelis2008}
B. Marcelis, S.J.J.M.F. Kokkelmans, G.V. Shlyapnikov, D.S. Petrov,
Phys. Rev. A {\bf 77}, 032707 (2008)

\bibitem{BlumePrivate}
D. Blume, private communication

\bibitem{note:addlargerstar}
The second term in Eq.~(\ref{eq:addlimit}) can be cast in the form
of an integral. Its analytic integration is not possible, but
numerically the result can be obtained with a very well controlled
accuracy.

\bibitem{Efimov1970}
V.N. Efimov, Yad. Fiz. {\bf 12}, 1080
(1970) [Sov. J. Nucl. Phys. {\bf 12}, 589 (1971)]

\bibitem{NielsenPhysRep2001}
E. Nielsen, D.V. Fedorov, A.S. Jensen, E. Garrido, Phys. Rep. {\bf 347}, 373 (2001)

\bibitem{braaten07} E. Braaten, H.-W. Hammer, Annals Phys. {\bf 322}, 120 (2007)

\bibitem{IncaoEsry2005}
J.P.~D'Incao, B.D.~Esry, Phys. Rev. Lett. {\bf 94}, 213201 (2005)

\bibitem{remark} In Ref. \cite{WangIncaoEsry2009} the atom-dimer relaxation in $s$-wave
  collisions in the case $m_\up=m_\down$ is considered by using a
  different method. Equation~(\ref{SWaveBareRateConst}) disagrees with
  the power law stated in that article. We think that the authors do
  not reach sufficiently large values of $R^*/a$.

\bibitem{WangIncaoEsry2009}
Y.~Wang, J.P.~D'Incao, B.D.~Esry, arXiv:0906.5019 (2009)

\bibitem{Blume}
J.~von~Stecher, C.~H.~Greene, D.~Blume, Phys. Rev. A {\bf 77}, 043619 (2008)

\bibitem{Nascimbene2009}
S.~Nascimbene, N. Navon, K. Jiang, L. Tarruell, M. Teichmann,
J. Mckeever, F. Chevy, C. Salomon, Phys. Rev. Lett. {\bf 103},  170402
(2009)

\bibitem{Mathy2010}
C.J.M.~Mathy, M.M.~Parish, D.A.~Huse, arxiv:1002.0101 (2010)

\bibitem{Radzihovsky2009}
L.~Radzihovsky, S.~Choi, Phys. Rev. Lett. {\bf 103}, 095302
(2009)

\bibitem{Zvonarev2007}
M.B.~Zvonarev, V.V.~Cheianov, T.~Giamarchi, Phys. Rev. Lett. {\bf 99},
240404 (2007)


\bibitem{Tan2008}
Y. Nishida, S. Tan, Phys. Rev. Lett. {\bf 101}, 170401 (2008)

\bibitem{Tan2009}
Y. Nishida, S. Tan, Phys. Rev. A {\bf 79}, 060701(R) (2009)

\end{thebibliography}

\end{document}